%% file: main.tex
\documentclass[sigconf,anonymous=false]{acmart}

\usepackage[scientific-notation=true]{siunitx}
\usepackage{amssymb}
\usepackage{amsmath}
\usepackage{multirow}
\usepackage{comment}
\usepackage{balance}
\usepackage[utf8]{inputenc}
\usepackage{url}

\newcommand{\anontext}[2]{#2} 

\begin{document}

\title{Comparing Respiratory Monitoring Performance of\\ Commercial Wireless Devices}

\input{authors}

\input{abstract}

\fancyhead{}
\settopmatter{printacmref=false} 
\renewcommand\footnotetextcopyrightpermission[1]{} 

\maketitle


\input{intro}

\input{related_methods}
\input{equipment}

\input{clinical_study}

\input{our_methods}

\input{results}

\input{future_conclusion}
\input{ack}

\balance
\bibliographystyle{abbrv}
\bibliography{refs}
\end{document}

%% file: authors.tex
%
%

\author{Peter Hillyard}
\affiliation{%
  \institution{Xandem Technology}
}
\email{peter@xandem.com}

\author{Anh Luong}
\affiliation{%
  \institution{Carnegie Mellon University}
}
\email{anhluong@cmu.edu}

\author{Alemayehu Solomon Abrar}
\affiliation{%
  \institution{University of Utah}
}
\email{aleksol.abrar@utah.edu}

\author{Neal Patwari}
\affiliation{%
  \institution{University of Utah \\ and Xandem Technology}
}
\email{npatwari@ece.utah.edu}

\author{Krishna Sundar}
\affiliation{%
  \institution{Health Sciences Center \\ University of Utah,}
}
\email{krishna.sundar@hsc.utah.edu}

\author{Robert Farney}
\affiliation{%
  \institution{Health Sciences Center \\ University of Utah,}
}
\email{robert.farney@hsc.utah.edu}

\author{Jason Burch}
\affiliation{%
  \institution{Health Sciences Center \\ University of Utah,}
}
\email{jason.burch@hsc.utah.edu}

\author{Christina A. Porucznik}
\affiliation{%
  \institution{Department of Family \\ and Preventive Medicine, \\ University of Utah \\ School of Medicine,}
}
\email{christy.porucznik@utah.edu}

\author{Sarah Hatch Pollard}
\affiliation{%
  \institution{Department of Surgery, \\ University of Utah \\ School of Medicine,}
}
\email{sarah.pollard@hsc.utah.edu}

%% file: abstract.tex
%
%
\begin{abstract}
This paper addresses the performance of systems which use commercial wireless devices to make bistatic RF channel measurements for non-contact respiration sensing.  Published research has typically presented results from short controlled experiments on one system. In this paper, we deploy an extensive real-world comparative human subject study. We observe twenty patients during their overnight sleep (a total of 160 hours), during which contact sensors record ground-truth breathing data, patient position is recorded, and four different RF breathing monitoring systems simultaneously record measurements.  We evaluate published methods and algorithms.  We find that WiFi channel state information measurements provide the most robust respiratory rate estimates of the four RF systems tested.  However, all four RF systems have periods during which RF-based breathing estimates are not reliable.
\end{abstract}

%% file: intro.tex
\section{Introduction}
In both in-patient and in-home health care settings, respiratory monitoring plays an important role in prognosis, diagnosis, and prevention of respiratory events, disease, and death. 
Sensing devices used to measure respiration are typically contact-based and are wired to a monitor. Wearing a sensor can limit mobility, interrupt daily activities or disrupt sleep. There is also a risk of the sensor becoming detached. Furthermore, for patients with sensitive skin (e.g., burn patients), applying, wearing, or removing a sensor may cause significant discomfort. 

Many non-contact respiration monitors have been developed to address the drawbacks of contact-based sensors. Included in this group are RF systems like Doppler and pulse radars \cite{droitcour2009respiratory}, WiFi devices which measure channel state information (CSI) \cite{liu2014wisleep}, and narrowband wireless devices which measure received signal strength (RSS) \cite{patwari2014breathfinding}. It has been shown that even the small displacement of a person's chest during respiration can change the magnitude or phase of these RF channel measurements. 

There is particular interest in channel measurements from WiFi, narrowband, and ultra-wideband (UWB) devices, devices which are becoming more ubiquitous in homes and businesses with the growth of the internet of things (IoT). These devices are already being used for wireless transfer of data between devices for sensing and automation services. As these devices are transmitting data, their channel measurements can simultaneously be used for respiration monitoring. 

Respiratory monitoring with WiFi, narrowband, and UWB devices have all been studied and evaluated individually. We survey RF-based, non-contact respiration monitoring. We compare measurements and devices, as well as methods for selecting the best channels from multi-channel measurements, filtering noise, detecting motion, and estimating respiration rate. 
One common limitation is that the experimental setup heavily influences the evaluation results, and these methods have not been compared to each other. Thus it is not clear how different monitoring systems perform in comparison to another.

In this paper, we provide for the first time a side-by-side comparison of the performance of four of these RF technologies in a real-world patient study.  During twenty overnight sleep studies of volunteer patients, we simultaneously measure the RF channel: the channel impulse response with a pair of UWB transceivers, channel state information with a pair of WiFi devices, and 1-dB quantized and sub-dB quantized RSS with pairs of narrowband devices. The abilities of each RF technology to estimate respiratory rate (RR) are compared to each other, as well as the performance as a function of position and motion of the patient.

The value of the side-by-side comparison with twenty overnight studies is increased due to the number of hours of data collected, the uncontrolled nature of the studies, and the fact that many patients have disordered sleep breathing events, {\it e.g.} apnea and hypopnea. We collect 160 hours of data during the course of the twenty studies. During each study, the patient sleeps in a bed in a room at a sleep clinic where they are free to sleep in a given position and to move in the bed at any time. These conditions provide a very realistic environment with which to compare the RF technologies as similar conditions will likely occur in a person's home in the same uncontrolled manner. This data is public and is made available at \anontext{[Citation removed for double-blind review]}{\cite{hillyard2018dataset}}. 

The data show that all four wireless devices can achieve as low as $0.24$ breaths per minute (bpm) median error during certain periods of time. We confirm that all wireless device fail to track the respiratory rate during other significant periods of time \cite{wang2016respiration}. During these failure periods, likely due to the particular arrangement of multipath components for a person's position, breathing can not be observed in the measurements. A failure period typically ends when the person moves. However, failure periods happen equally during different facing directions (supine, prone, left, and right). Particularly surprising was that even wireless devices using orders of magnitude higher bandwidth (CIR, CSI) and multiple antenna pairs (CSI) are unable to achieve high reliability. 

Overall, this study shows that WiFi CSI provides the most robust estimates of RR. In terms of algorithms, our experimental results show that estimating RR using the frequency response of the measurements outperforms methods that are based on the time-domain inter-breath interval.  

We emphasize that the contribution of this paper is to provide an extensive real-world experimental setup and carefully collected data set in which four RF breathing monitoring systems and ground-truth RR data can be compared side-by-side, to our knowledge, for the first time. We hope that the results can provide direction to an active area of research and influence future systems to achieve greater performance.














%% file: related_methods.tex
\section{Related Work} \label{S:related_methods}
RF-based respiration monitoring originated from observations that phase shifts of microwave chirp signals reflected off of a nearby, stationary person matched the person's breathing frequency \cite{lin1975respiration}. Since that time, there have been significant advancements in wireless, RF-based respiration monitoring. These advancements include the type of RF channel measurements used, the availability of multidimensional measurements, motion detection methods, and methods for estimating RR. In this section, we survey several different approaches that have been developed for respiration monitoring.

\subsection{Measurement Methods}
While the RF technology used to monitor respiration has evolved and expanded in capability over the last 40 years, they all leverage the same underlying physics. First, a transmitter sends a signal, and then that signal's amplitude and phase are modulated by the inhalation and exhalation of a person's chest wall. The modulated signal then arrives at a receiver which measures the changes caused by respiration. Differences in the technologies include the type of signal transmitted, the distance between the transmit and receive antenna(s), the number of transmitters and the number of receivers, and how the receiver measures the RF channel.

We categorize each technology as either a monostatic or multistatic device. Multistatic devices have one or more transmit and receive antennas separated by at least 30 cm whereas monostatic devices have just one transmit antenna and one or more receive antennas but which are no more than 30 cm apart. In this paper, we focus on multistatic devices, but point to a few monostatic devices like doppler radar \cite{nguyen2016breathing,graichen2012vital,dei2009breathing,li2010vital,chen1986lifedetection,ballal2012breathing,hu2014cardiopulmonary,zhao2017physiological,xia2016vital,li2016breathing,tu2016respiration,hu2014vital,salmi2012vital,li2010respiration,droitcour2009respiratory,droitcour2009noncontact,yang2017vital}, frequency-modulated continuous-wave radar/sonar \cite{peng2017vital,adib2015breathing,anitori2009lifesign,nandakumar2015apneaapp,nandakumar2017covertband}, pulse radar \cite{nguyen2013vital,wang2012respiration,lai2011respiratory,salmi2011estimation,leib2010vital,immoreev2008monitoring,chen2008respiration,ossberger2004respiratory}, and pulse doppler radar \cite{wisland2016vital,lazaro2016vital} for completeness.

{\bf Multistatic}:
In multistatic systems, a transmitter and receiver are placed so that the link line between them passes near the chest of a breathing person. 
The phase and amplitude of some of the multipath components of the transmitted signal are changed as the person's chest expands and contracts. The receiver makes a measurement of the RF channel which captures the changes in the multipath. 

Standard commercial wireless devices are separated in space as their objective is typically to transmit data across a distance.  When they are used to make channel measurements, these measurements are multistatic.  Thus while
monostatic devices were typically designed for sensing, multistatic device measurements are a new opportunity to repurpose transceivers that are low cost because of their ubiquitous use in commercial devices, and are often already in people's homes. This opens the opportunity to use existing infrastructure for sensing purposes like respiration monitoring. We focus our comparison in this paper on multistatic systems for the reasons above. We describe a few of these multistatic systems in this section.

UWB-IR is used in both monostatic and multistatic systems. We describe in Section \ref{S:preprocess} how the channel impulse response (CIR) of UWB devices is used in a multistatic system to monitor respiration. This method is also implemented in past research  \cite{venkatesh2005respiration,rivera2006respiration,baboli2009respiration,baboli2012respiration,kilic2014uwb}. While monostatic systems generally use the time of flight, multistatic systems make use  of the channel impulse response (CIR) which is measured from all of the time-delayed copies of the transmitted pulse that arrive at the receiver. The phase and magnitude of some time delay bins in the channel impulse response, which we call taps, will show changes as a person breathes \cite{venkatesh2005respiration,rivera2006respiration,baboli2009respiration,baboli2012respiration,kilic2014uwb}. These phase or magnitude measurements can thus be used to monitor the respiratory activity of a person. 

Modern WiFi routers use orthogonal frequency division multiplexing (OFDM) to combat frequency selective fading. Recent driver modifications have given access to complex-valued signals on many subcarriers called channel state information (CSI) at the PHY layer of a WiFi enabled device \cite{halperin2011csitool,xie2015wifi}. It has been shown that the magnitude and phase of the complex-value signal on many subcarriers \cite{wu2015detection,wang2017phasebeat,ravichandran2015wibreathe,liu2015vital,liu2016respiration,shang2016vital,ma2016respiration,wang2016respiration,liu2014wisleep,chen2017breathing} and the received signal strength (RSS) \cite{abdelnasser2015ubibreathe} are affected by the chest movements of a breathing person. The ubiquity of WiFi routers and devices in homes and buildings makes this technology an attractive means of performing respiration monitoring. WiFi devices commonly have multiple antennas which offer additional MIMO links. Each MIMO link adds an additional set of subcarriers at the receiver that can be used for respiratory monitoring. For example, a $3 \times 3$ MIMO link was used in \cite{liu2016respiration}. The number of WiFi access points (transmitters) and the number of WiFi clients (receivers) varies from three transmitters and three receivers \cite{liu2015vital}, two transmitters and three receivers \cite{liu2016respiration,liu2014wisleep}, or just one transmitter and one receiver \cite{wu2015detection,wang2017phasebeat,ravichandran2015wibreathe,abdelnasser2015ubibreathe,shang2016vital,ma2016respiration}. More transmitters and receivers are used to get more spatially diverse measurements. Additionally, both the 2.4~\si{\giga\hertz} \cite{ravichandran2015wibreathe,abdelnasser2015ubibreathe,liu2015vital,liu2016respiration} and the 5~\si{\giga\hertz} \cite{wang2017phasebeat,ma2016respiration,chen2017breathing} bands have been used in CSI-based respiration monitoring systems.

It has been shown that when one or many Zigbee links are nearby a breathing person, the RSS value of the links change as a person inhales and exhales \cite{patwari2014breathing,patwari2014breathfinding,kaltiokallio2014respiration,zhao2016respiration,hostettler2017respiration}.  IEEE 802.15.4 transceivers often make RSS values available to the application. The Zigbee devices in these systems were programmed to measure RSS on 16 \cite{kaltiokallio2014respiration,hostettler2017respiration}, 4 \cite{zhao2016respiration,patwari2014breathfinding}, and 1 frequency channel \cite{patwari2014breathing} in the 2.4 GHz band. To increase spatial diversity, 33 \cite{patwari2014breathfinding}, 20 \cite{patwari2014breathing}, 4 \cite{zhao2016respiration}, and 1 \cite{kaltiokallio2014respiration} transceiver(s) are deployed to create a mesh network of sensing links. The purpose of increasing the number of transceivers and channels is to increase the number of links which may be sensitive to breathing, in a form of frequency diversity.

The RSS value on most IEEE 802.15 devices is quantized with 1 dB step sizes, limiting how sensitive a link is to the small displacement of a person's chest during breathing. An alternative system was developed to achieve sub-dB quantization step sizes \cite{luong2016stepsize} to provide greater sensitivity to breathing. The radio sends a CW signal at 434~\si{\mega\hertz} and uses one transmitter and one receiver.

\subsection{Stream Selection}
The multistatic systems described above commonly measure multidimensional signals. For CIR, the multidimensional signal is the magnitude or phase of each tap. For CSI, the multidimensional signal is the magnitude or phase of each subcarrier on each MIMO link. For RSS, the multidimensional signal is each channel on which received signal strength is measured. In this section, we refer to an individual tap, subcarrier, or channel in the multidimensional signal as a stream. When observed over time, each stream has the potential of showing the changes due to a person's respiration. It is common however for some streams to have a higher signal-to-noise ratio (SNR) than other streams. For reliable respiration monitoring, selecting the best stream(s) is necessary. Different methods of picking the ``best'' stream or streams have been developed. 

Several works find that streams with the highest variance \cite{baboli2012respiration,baboli2009respiration} or average squared value \cite{venkatesh2005respiration,rivera2006respiration} are those that are most sensitive to breathing. This comes from observations that some streams are periodic with regular breathing and had high peak-to-peak values compared to other streams. The streams with higher peak to peak values would also have higher variances or average squared values. A more computationally intensive method was proposed where the mean absolute deviation was computed for each stream \cite{wang2017phasebeat}. The middle of the streams with the greatest mean absolute deviations was selected. This method filters out streams with very low peak to peak values, but then also filters out streams that may have very high peak to peak values that may be associated with noise rather than the influence of respiration.

In other methods, the best stream was selected based on how periodic the signal was. Periodicity was measured by the ratio of the stream's amplitude to the RMSE between the stream measurement and a single frequency sinusoid \cite{liu2016respiration} or from a recurrence plot \cite{liu2014wisleep}. Selecting a stream based on periodicity is done in hopes that the more periodic the signal, the better the performance of the respiration monitoring.

When many streams have been measured, it may not make sense to pick only one. There may be several streams that are ``best'' and using all of the best streams could help average out noise. For example, in \cite{liu2015vital}, each stream is weighted by its variance. Streams with higher variances are given greater weights. The idea of weighting streams is also used in \cite{patwari2014breathfinding,patwari2014breathfinding,kaltiokallio2014respiration,zhao2016respiration} but all streams are given equal weight. The idea is that, if most of the streams show are affected by a person's respiration, then their joint contribution will be better able to monitor respiration.

\subsection{Motion Detection and Removal}
During respiration monitoring, RF devices measure very small displacements of a person's chest during respiration. Larger motion like walking, moving an arm or leg, or even muscle twitches can induce very large changes in RF measurements. During periods of motion, it is very difficult to recover the breathing signal as it is overwhelmed by effects of motion. Many motion detection algorithms have been developed to flag RF measurements as happening during motion events. We describe of few of these detectors here.

Several detectors detect when the moving variance exceeds a threshold \cite{baboli2009respiration,wang2017phasebeat,liu2015vital}. For multidimensional channel measurements, motion is detected when the average of the moving variances exceeds a threshold \cite{liu2015vital}. In another detector, the residual of least-squares harmonics method is use to detect motion \cite{ravichandran2015wibreathe}. When a window of signals has only breathing, the residual will be low since it closely matches a single-frequency sinusoid. During motion, the residual is large since many frequencies are in the measurements during motion. Another motion detector leverages the idea that channel measurements made during motion are much less periodic during motion than during respiration \cite{adib2015breathing}. This method is described in more detail in Section \ref{S:motion_detection}. A threshold is placed on the ratio of the maximum amplitude of a fast Fourier transform (FFT) to the average amplitude of the non-maximum amplitudes. The greater the difference between the maximum amplitude and the average amplitude, the greater the evidence that the signal is periodic and thus from a time period of motionless and regular respiration. The spectral content during motion, on the other hand, appears more flat.

Some motion detectors collect training data where motion and no motion measurements are made. One of these motion detectors finds a linear discriminant using an SVM where the maximum eigenvalue of both the magnitude and phase are the features for the SVM classifier \cite{wu2015detection}. The maximum eigenvalues are generally high during no motion, and low during motion. In a second motion detector, coefficients from a continuous wavelet transform and stationary wavelet transform, and the median absolute deviation are used as features for a decision tree \cite{graichen2012vital}. A two-state hidden Markov model was also used where the conditional distribution during motion had a higher variance than in the no motion conditional distribution \cite{kaltiokallio2014respiration}. 

Other motion detection methods remove the affect of motion in the signal. This has been accomplished with breakpoint detection. When a breakpoint is detected, the mean is computed and removed in the window of samples before and after the breakpoint \cite{abdelnasser2015ubibreathe,patwari2014breathfinding,liu2016respiration}. The way breakpoints are detected vary from Wilcox U test \cite{patwari2014breathfinding} to Kolmogorov-Smirnov test \cite{liu2016respiration}.

\subsection{Respiratory Rate (RR) Estimation}
A common way to evaluate the performance of a respiration monitoring system is to estimate the RR of a person and compare the estimate to ground truth. Various methods have been developed to estimate the RR. We describe a few of these methods here.

In that a person's respiration has a sinusoidal appearance during regular breathing, many methods window 10 - 30~s of channel measurements and estimate the spectral content \cite{wisland2016vital,lazaro2016vital,lai2011respiratory,jia2013lifesign,xia2016vital,tu2016respiration,leib2010vital,peng2017vital,salmi2012vital,li2010respiration,venkatesh2005respiration,baboli2009respiration,abdelnasser2015ubibreathe}. The assumption is that, during that window of time, there is very little change in the durations of inhalation and exhalation. Under this assumption, the magnitude of the frequency response will be maximum at the frequency of respiration. Other methods of estimating the frequency content of the window include finding the frequency at which a pure sinusoid at a given frequency and the window of samples most agree. This can be done with with a maximum likelihood estimate solution \cite{patwari2014breathfinding,patwari2014breathing,kaltiokallio2014respiration,zhao2016respiration,chen2008respiration}, least squares \cite{ballal2012breathing,rivera2006respiration}, or an iterative method which converges on a solution \cite{salmi2011estimation,li2010vital,wu2015detection}. Examining the spectral content of the signal is by far the most commonly used approach to estimate RR.

Alternatively, other methods compute the inter-breath interval (IBI) by detecting peaks in the measurement \cite{wang2012respiration,hu2014cardiopulmonary,hu2014vital,ravichandran2015wibreathe,liu2015vital,shang2016vital}. The time between peaks is saved for a window of 10 - 30~s which gives the respiratory period. The RR is the inverse of the IBI period. Other RR estimation methods include Kalman filtering a periodic Gaussian process  \cite{hostettler2017respiration}, and an iterative root-MUSIC approach \cite{chen2017breathing}.
The average of the respiratory rates during the last window of time is then used as the estimated respiratory rate.

Whereas some methods select the best channel to perform RR estimation, other methods have been developed to estimate an RR for each stream and then pick the ``best'' RR estimate. For example, an RR is estimated for each stream, outlier RR estimates are filtered out, and the average of the remaining RR estimates can be used as the final RR estimate \cite{wu2015detection}. Alternatively, each stream can be weighted based on its variance or another metric. A weighted average of the respiratory rates from each stream could be considered as the best estimate \cite{liu2015vital}. Alternatively, instead of first estimating the respiratory rate, the magnitude of frequency response of each stream could be averaged together to get an average power spectral density \cite{patwari2014breathfinding,patwari2014breathing,kaltiokallio2014respiration,zhao2016respiration,chen2008respiration}. The RR is then the frequency at which the average power spectral density is maximum.

\subsection{Filtering Channel Measurements}
It is common practice for channel measurements to be filtered prior to respiration rate estimation. Low-pass filters are used to attenuate frequencies higher than a person is likely to breathe. Cutoff frequencies are reported as low as 0.4~\si{\hertz} (24 bpm) and as high as 1~\si{\hertz} (60 bpm). For peak finding methods, polynomial fitting may also be used to suppress false peaks \cite{shang2016vital}. 

Methods that estimate spectral content typically apply mean removal. A high-pass filter with a low cutoff frequency is used \cite{nguyen2013vital,leib2010vital,adib2015breathing,jia2013lifesign,anitori2009lifesign,yang2017vital,tu2016respiration,salmi2012vital,li2010respiration,droitcour2009respiratory,ravichandran2015wibreathe,abdelnasser2015ubibreathe,patwari2014breathing,kaltiokallio2014respiration}, or the mean of a window of measurements is subtracted from each new measurement \cite{venkatesh2005respiration,lazaro2016vital,nguyen2016breathing,ma2016respiration,zhao2016respiration}. For CSI measurements, median filters are used to filter out heavy-tailed noise \cite{liu2015vital,liu2016respiration,shang2016vital,liu2014wisleep}.

%% file: equipment.tex
\section{Equipment} \label{S:equipment}
In this section, we describe the testbed used to study four different RF-based respiratory monitoring systems. We also describe the polysomnography equipment used to collect ground truth data.

\subsection{RF System Testbed}
\begin{figure}
\begin{center}
\includegraphics[width=0.75\columnwidth]{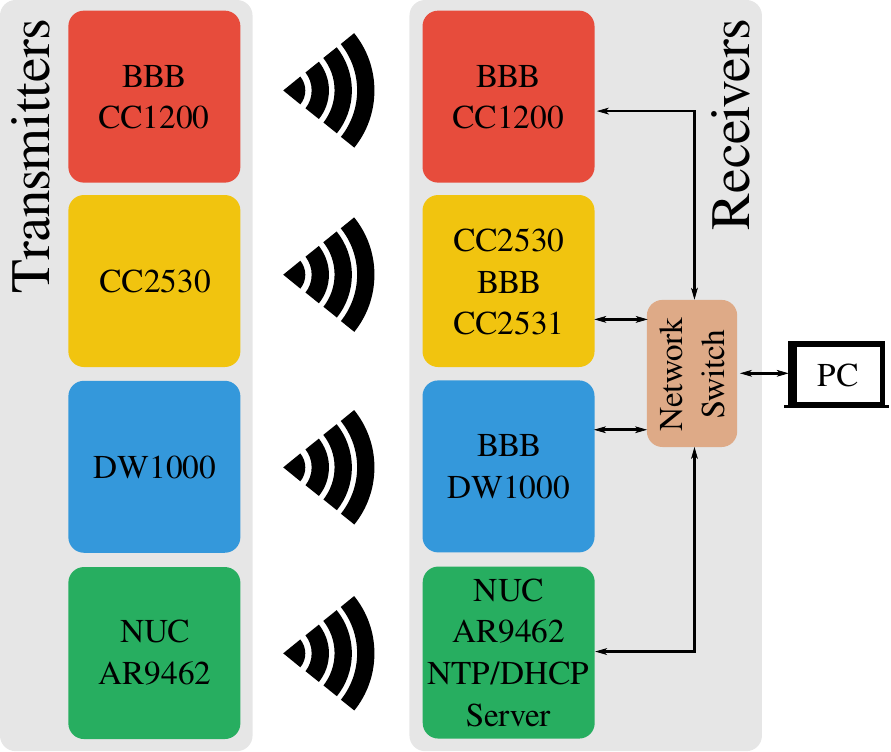}
\end{center}
\caption{The components of the RF testbed. The transmitter components are contained in the box on the left, and the receiver components are contained in the box on the right.}
\label{F:overview}
\vspace{-1em}
\end{figure}

In this section, we discuss the design of the four different systems representing the current state-of-the-art in non-contact multistatic RF respiratory monitoring including UWB-IR, WiFi CSI, Zigbee RSS, and sub-1 dB quantized RSS. The components of this system are discussed in the following sections and are shown in Fig.~\ref{F:overview}.

{\bf Sub-dB RSS}:
A CC1200 radio with a WA5VJB Log Periodic 400-1000~\si{\mega\hertz} antenna is placed in both the transmitter and receiver box. The CC1200 transmitter and receiver are controlled with a BeagleBone Black (BBB). The transmitter sends a 900~\si{\mega\hertz} continuous wave and the receiver measures sub-1 dB quantized RSS measurements as described in \cite{luong2016stepsize}. The measurements are stored on the receiver's BeagleBone Black. We refer to this measurement system as SUB.


{\bf Zigbee RSS}:
One CC2530 transceiver with a WA5VJB Log Periodic 900-2600~\si{\mega\hertz} antenna is placed in each of the transmitter and receiver boxes. A logging CC2531 radio is attached to a BBB to save the RSS measured between the two transceivers. The transceivers use TDMA to take turns transmitting while looping through all sixteen \SI{2.4}{\giga\hertz} Zigbee channels. We refer to this measurement system as RSS. We will distinguish between the system and the measurement when necessary. 


{\bf WiFi CSI}:
We replace the existing WiFi card in an Intel NUC D54250WYK with an Atheros AR9462. We use the CSI tool developed in \cite{xie2015wifi} and modify the kernel driver to operate in the WiFi ~\SI{5}{\giga\hertz} band. Two WA5VJB Log Periodic 2.11-11.0~\si{\giga\hertz} antennas are attached to the WiFi card to enable $2\times2$ MIMO. One modified Intel NUC serves as the access point and is placed in the transmitter box. Another modified NUC serves as the client and is placed in the receiver box. The client pings the access point and records CSI for 114 subcarriers on each MIMO link. We refer to this measurement system as CSI. We will distinguish between the system and the measurement when necessary.


{\bf UWB-IR}:
A Decawave EVB1000 is placed in the transmitter box. The transmitter sends UWB packets on a channel that occupies 3.77 - 4.24~\si{\giga\hertz}. A second EVB1000 in the receiver box measures the CIR and sends the complex-valued CIR taps to a BBB. Both the transmitter and receiver use the PCB UWB antenna provided with the EVB1000. We refer to this measurement system as CIR. We will distinguish between the system and the measurement when necessary.

\begin{figure}
  \begin{center}
  \includegraphics[width=0.75\columnwidth]{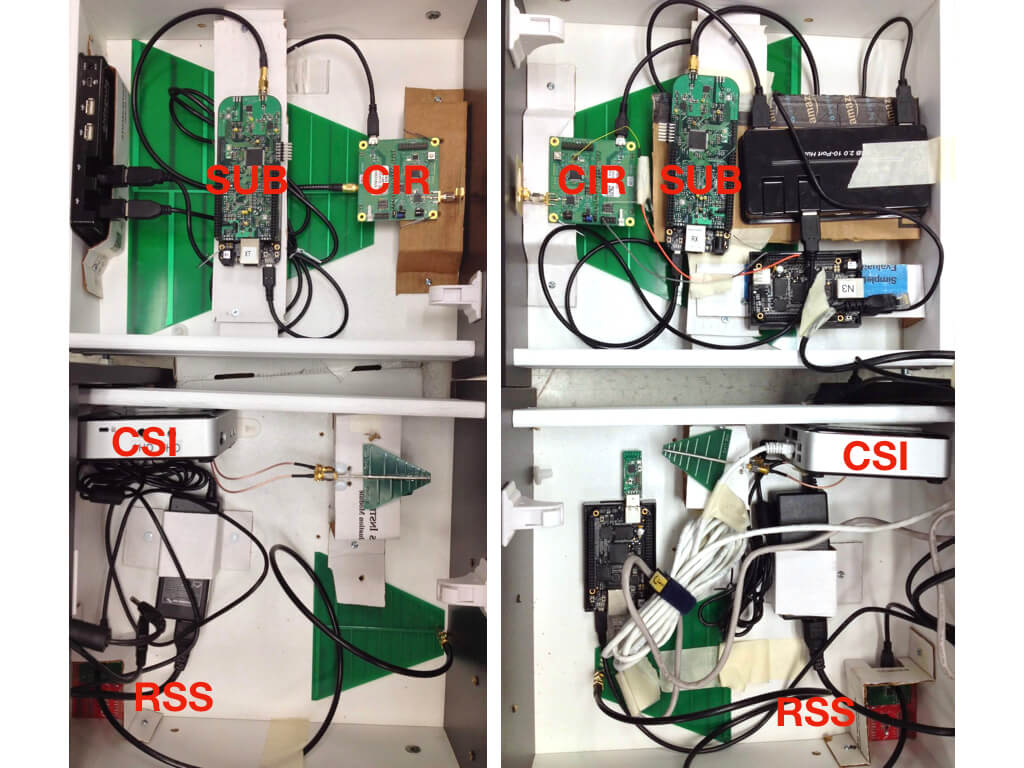}
  \end{center}
  \caption{RF transmitters (left) and receivers (right) enclosed in the drawers of a bedside dresser.}
  \label{F:rf_boxes}
  \vspace{-1em}
\end{figure}

{\bf Network}:
The RF devices in the receiver box are attached to the a NetGear 5-port switch and are time synchronized using NTP with the Intel NUC as the NTP server. The Intel NUC is also a DHCP server. 
The devices are housed in separate boxes (see Fig.~\ref{F:rf_boxes}).

{\bf Polysomnography}:
Patients who come for a sleep study are dressed with a number of sensors including respiratory impedance plethysmography (RIP) belts around the chest and abdomen, and a thermistor and nasal cannula sensor in their nose. These sensors are plugged into an amplifier, and their measurements are read into Natus SleepWorks Software \cite{natus2017sleepworks} for visualization. The data collected are exported as an EDF and converted to ASCII \cite{vanbeelen2017edf} to be processed offline.


%% file: clinical_study.tex
\section{Clinical Study} \label{S:clinical_study}
\begin{figure}
  \begin{center}
  \includegraphics[width=0.80\columnwidth]{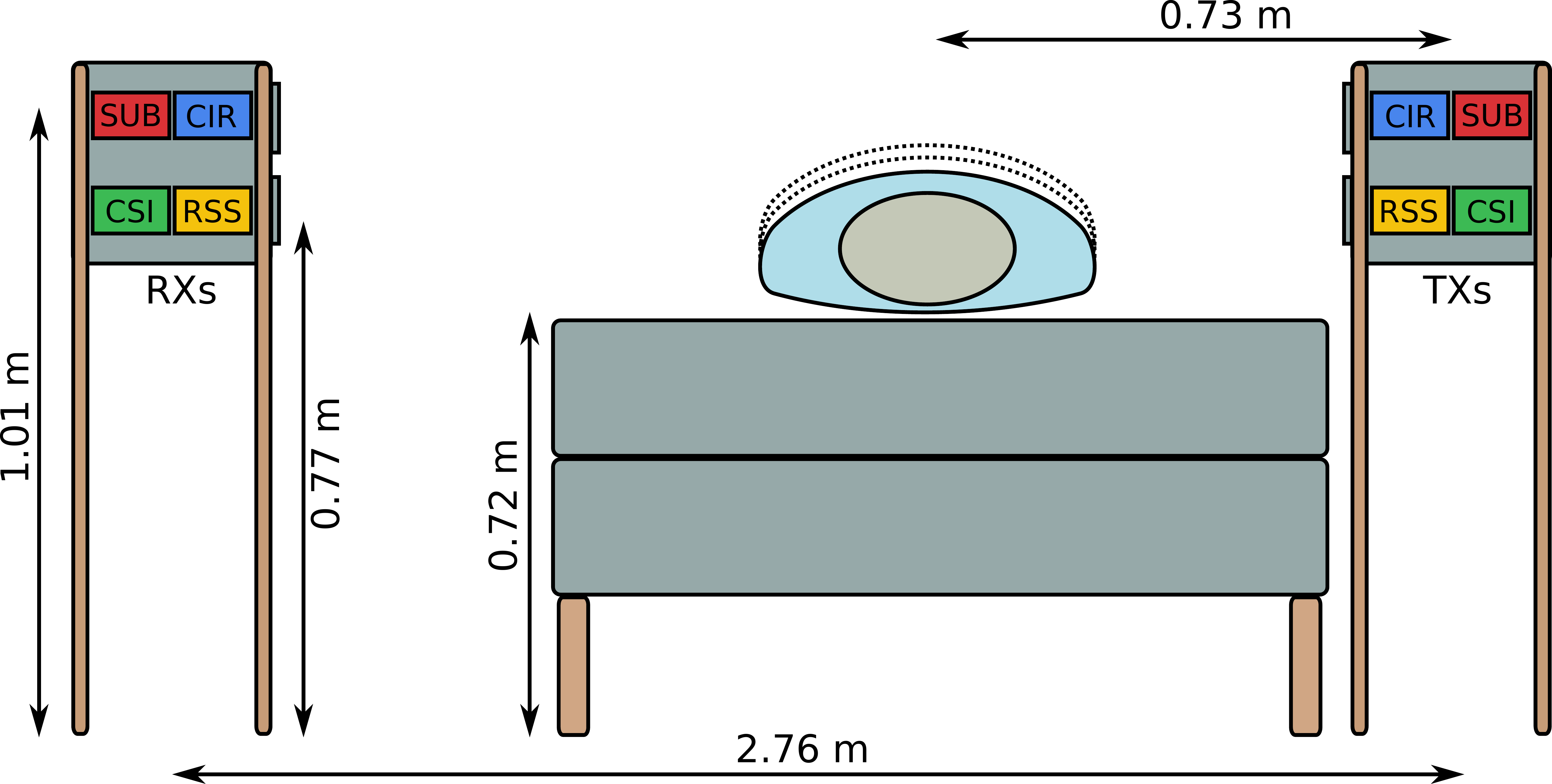}
  \caption{Position of the RF sensor and the patient's bed during each clinical study. Relevant heights and distances are included.}
  \label{F:testbed_location}
  \end{center}
  \vspace{-1em}
\end{figure}

In our clinical study, 20 patients, who were already scheduled for a regular 8 h sleep study, were asked to participate in a breathing monitoring experiment. Willing participants read and signed a consent form for IRB \anontext{XXX [removed for double-blind review]}{00084836}. Thereafter, the RF testbed was turned on and then positioned so that the link line between transmitter and receiver was perpendicular to and on top of the person's chest as shown in Fig.~\ref{F:testbed_location}.
The patient was then outfitted with polysomnograph sensors and, once in bed, the sleep study began. During the study, registered polysomnographic technologists annotated the polysomnograph data with the time and duration of pertinent events related to sleep. These annotations were reviewed a second and third time by other technicians and physicians.

The events recorded by the sleep technicians include limb movements, arousals, obstructive hypopnea and apnea, central apnea, sleep stage, and sleeping position. While these events are important for rating sleep quality and diagnosing sleep disorders, we do not make use of them in this paper. The polysomnography and RF data and annotated events from all twenty studies have been anonymized and made publicly available for future researchers' use \anontext{[Citation removed for double-blind review]}{\cite{hillyard2018dataset}}. %

Of the twenty patients, eleven were male and nine were female. The median age of the males was 55 years old and 60 for females. The median height and weight for females was 1.65~\si{\meter} and 99.79~\si{\kilo\gram} and for men 1.72~\si{\meter} and 88.45~\si{\kilo\gram}.

%% file: our_methods.tex
\section{Methods} \label{S:Methods}
\begin{figure}
  \begin{center}
  \includegraphics[width=0.9\columnwidth]{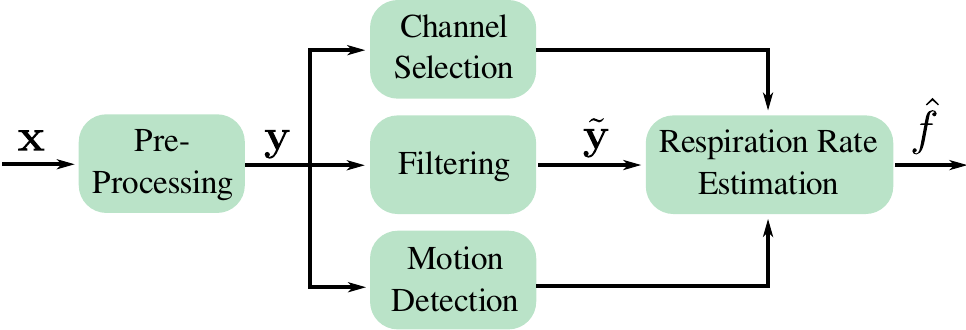}
  \caption{Breathing monitoring blocks to perform signal processing on RF measurements, to select the best streams from a multidimensional signal, to estimate respiration rate $\hat{f}$, and to detect motion periods.}
  \label{F:breathing_monitor_system}
  \end{center}
  \vspace{-1em}
\end{figure}

In this section, we evaluate a variety of methods that are commonly used in RR estimation. These methods can be categorized into the blocks shown in Fig.~\ref{F:breathing_monitor_system}. The multidimensional channel measurement $\mathbf{x}$ is fed into a pre-processing block. A series of filters in the filter block attenuates undesired frequency content. Streams of the multidimensional measurements are then selected to be used in respiratory rate estimation. A motion detection block is used to ignore RR estimates during motion. The methods that we evaluate are either previously published methods, or adapted methods that we develop in this paper.

%
%
\subsection{Pre-Processing} \label{S:preprocess}
Each RF technology requires unique signal processing algorithms for each RF system in order to extract a breathing signal, as described in this section.

\begin{figure}
  \begin{center}
  \includegraphics[scale=0.75]{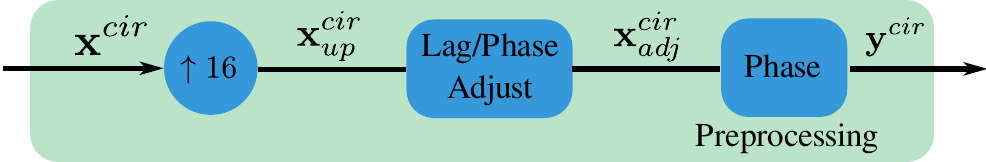}
  \caption{Pre-processing blocks for CIR.}
  \label{F:cir_preproc}
  \end{center}
  \vspace{-1em}
\end{figure}

{\bf Channel Impulse Response}:
The pre-processing block for a CIR measurement is shown in Fig.~\ref{F:cir_preproc}.
 First, a complex-valued CIR measurement is sent from the Decawave RX to a BBB at a sampling rate of $f_s^{cir} = 18.9$~\si{\hertz}. The channel impulse response measurement, $\mathbf{x}^{cir}_{raw} \in \mathbb{C}^{20}$, includes five complex-valued taps before and fifteen complex-valued taps after the first arriving path where each tap is a \SI{1}{\nano\second} bin. To provide greater time-delay resolution of arriving multipath, we upsample $\mathbf{x}^{cir}_{raw}$ by $N_{up}=16$ to get $\mathbf{x}^{cir}_{up}$.

\begin{figure}
  \begin{center}
  $\vcenter{\hbox{\includegraphics[scale=0.22]{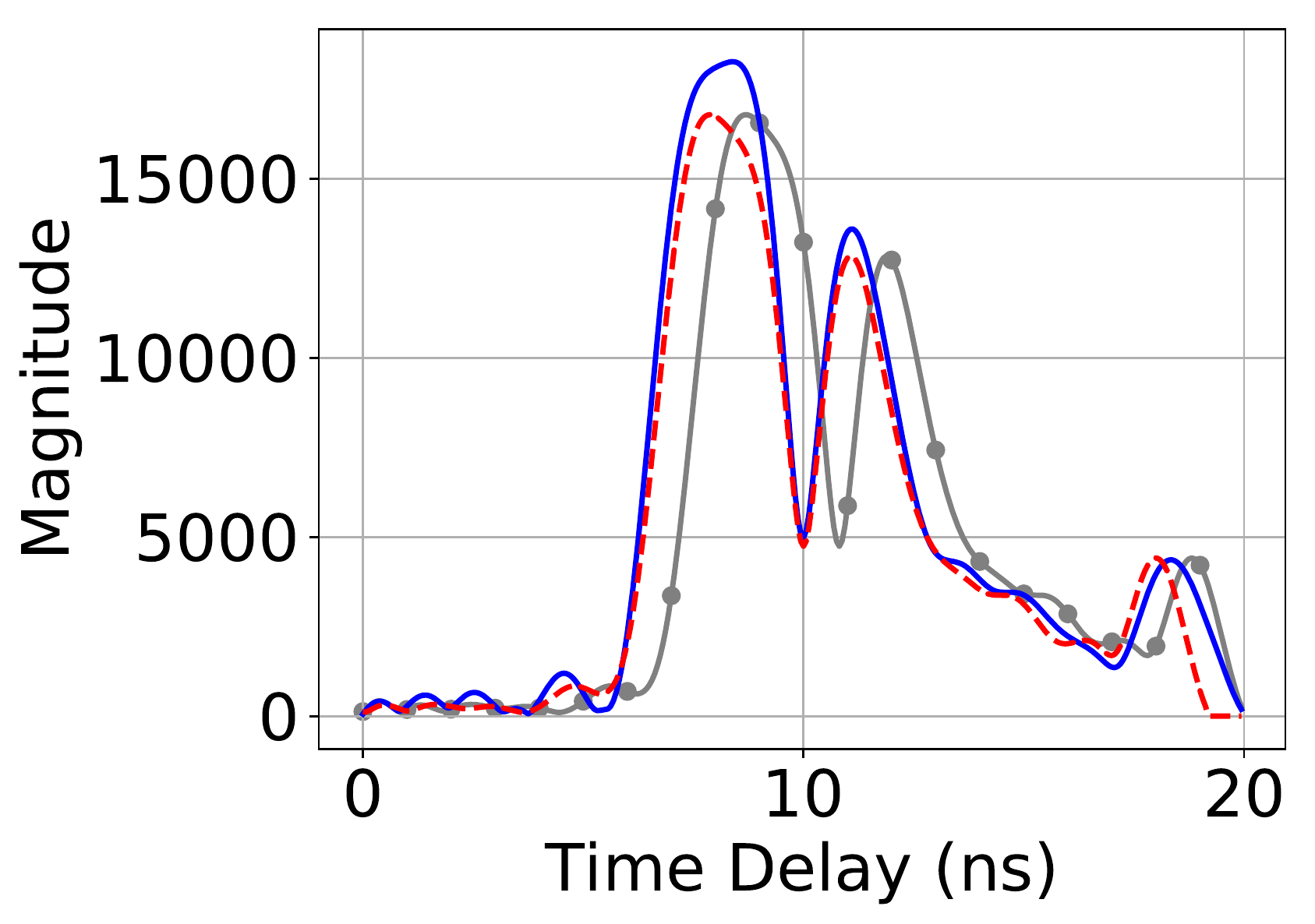}}}$\quad
  $\vcenter{\hbox{\includegraphics[scale=0.22]{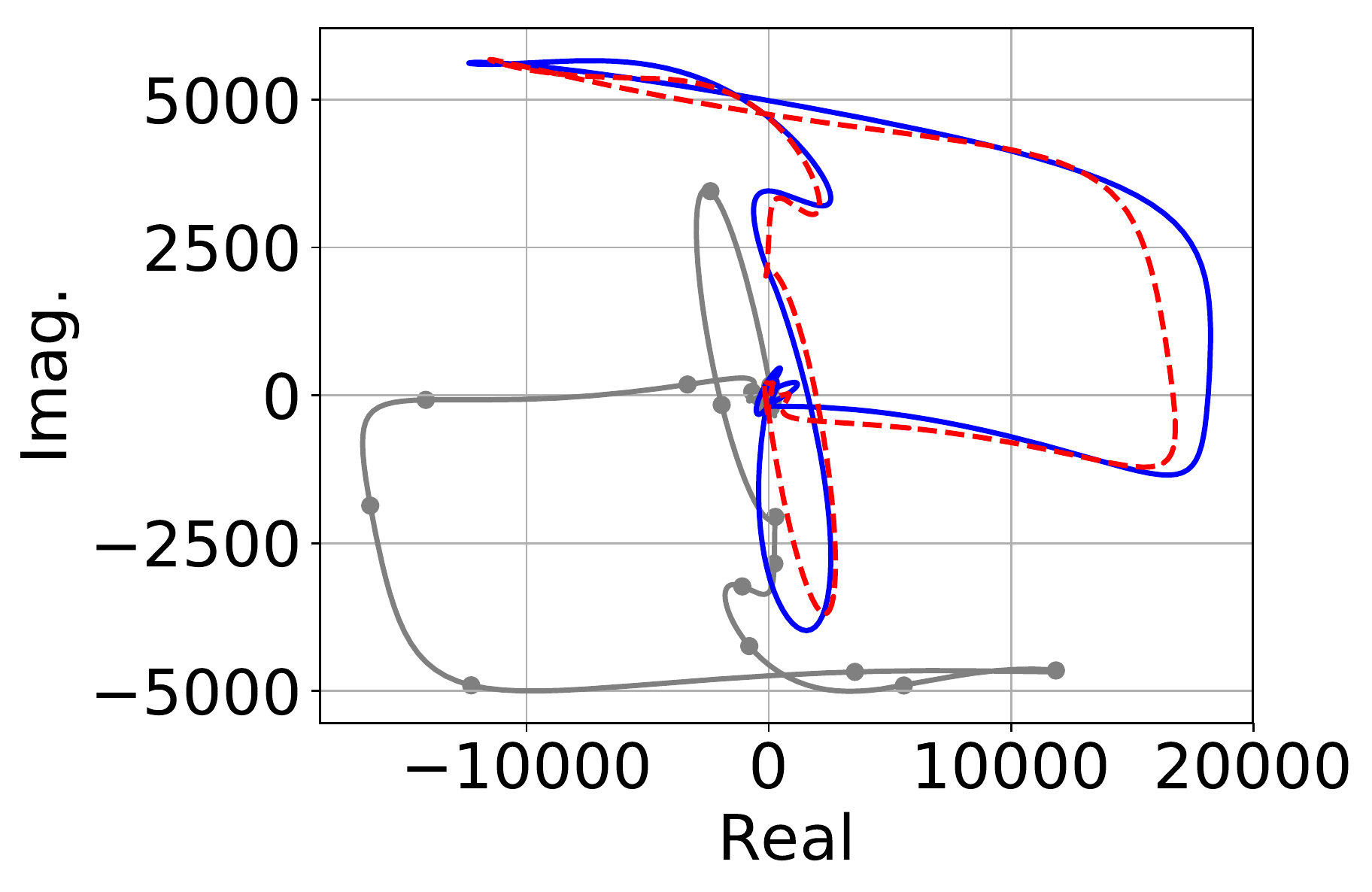}}}$
  \caption{The (left) magnitude and (right) complex-valued channel impulse response measurement before (gray) and after (dashed red) lag and phase adjustment with reference to a running average reference (blue).}
  \label{F:lag_phase_adjust}
  \end{center}
  \vspace{-1em}
\end{figure}

The UWB transmitter and receiver can both introduce lag and a phase rotation between measurements. This needs to be corrected to extract useful breathing signals. To correct for lag, we compute the cross-correlation of $\| \mathbf{x}^{cir}_{up} \|$ with the magnitude of a reference CIR. The lag is computed by finding the delay where the maximum correlation occurs. The reference CIR, $\mathbf{x}^{cir}_{ref}$, is an exponentially weighted average of the upsampled, lag and phase corrected measurements which is updated with every new CIR measurement. We obtain $\mathbf{x}^{cir}_{no}$ after $\mathbf{x}^{cir}_{up}$ is shifted by the computed delay to correct for lag. Complex zeros fill the vacant positions. Next, we rotate its phase to match the reference CIR:
\begin{equation}
\mathbf{x}^{cir}_{adj} = e^{j\hat{\theta}} \mathbf{x}^{cir}_{no}, \mbox{ for } \hat{\theta} = \arg \min_\theta \big\| \mathbf{x}^{cir}_{ref} - e^{j\theta} \mathbf{x}^{cir}_{no} \big\|
\end{equation}
Fig.~\ref{F:lag_phase_adjust} shows an example of a lag and phase adjusted CIR.
 In our own testing, we found that the phase of $\mathbf{x}^{cir}_{adj}$ captured the breathing signal better than the magnitude. The output of the pre-processing block for CIR measurements is $\mathbf{y}^{cir} = \angle \mathbf{x}^{cir}_{adj}$.

\begin{figure}
  \begin{center}
  \includegraphics[width=0.55\columnwidth]{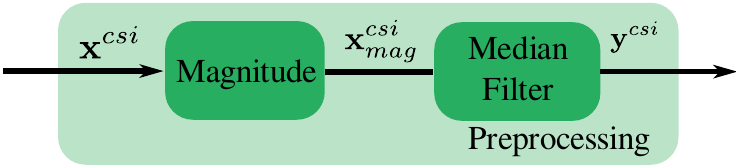}
  \caption{Pre-processing blocks for CSI.}
  \label{F:csi_preproc}
  \end{center}
  \vspace{-1em}
\end{figure}

{\bf Channel State Information}:
The signal processing blocks for a CSI measurement are shown in Fig.~\ref{F:csi_preproc}.
 A multidimensional complex-valued CSI measurement $\mathbf{x}^{csi}_{raw}$ from $2 \times 2$ MIMO is saved at a sampling rate of $f_s^{csi} = 9.9$~\si{\hertz}. 114 subcarriers are measured on each antenna pair and so  $\mathbf{x}^{csi}_{raw} \in \mathbb{C}^{456}$. We compute $\mathbf{x}^{csi}_{mag} = 10\log_{10}\mathbf{x}^{csi}_{raw}$. Heavy tailed noise is present in $\mathbf{x}^{csi}_{mag}$ which adds in a high-frequency component into the breathing signal which we wish to remove. We apply a median filter over a 0.7~\si{\second} window to $\mathbf{x}^{csi}_{mag}$ to remove this noise. The output of the pre-processing block for CSI measurements is $\mathbf{y}^{csi}$.

{\bf Sub-dB Quantized RSS}:
The CC1200 provides one power measurement $\mathbf{x}^{sub}_{raw} \in \mathbb{R}^{1}$ at a sampling rate of $f_s^{sub} = 487.5$~\si{\hertz}. We downsample by 30 by taking the average of 30-sample chunks. We compute the $\mathbf{x}^{sub}_{mag} = 10\log_{10}\mathbf{x}^{sub}_{raw}$ to get the sub 1-dB quantization RSS. These measurements are also corrupted with heavy tailed noise, and so we apply a median filter over a 0.45~\si{\second} window to $\mathbf{x}^{sub}_{mag}$ to remove large spikes. The output of the pre-processing block for SUB is $\mathbf{y}^{sub}$.

{\bf 1-dB Quantized RSS}:
Unlike the other RF measurements, RSS measurements do not need any special pre-processing. The RSS measurements are saved at a sampling rate of $f_s^{rss} = 4.5$~\si{\hertz}. At this stage, we refer to the RSS measurements as $\mathbf{y}^{rss} \in \mathbb{Z}^{32}$.

%
%
\subsection{Stream Selection} \label{S:stream_select}
Destructive and constructive interference of multipath components results in some streams being more sensitive to respiration than others. Stream selection is used to either remove unwanted streams, or weight the stream based on some metric for RR estimation. The stream selection algorithms presented in previous research do not perform well with the data we collected. As such, we present adapted stream selection algorithms for each RF channel measurement.

{\bf Channel Impulse Response}:
From our testing, we found that some of the streams of some elements in $\mathbf{y}^{cir}$ had high noise levels, resulting in a high variance signal that overwhelmed any possible breathing signal. RR estimates degrade when these streams are included in the estimation and detection algorithms. We filter out these streams by first computing a moving variance of the phase of each stream over a 30~\si{\second} window. With each new measurement, we filter out the streams whose variance is greater than the 50th percentile of the variances. 
It is possible that unwrapping the phase could increase the respiratory sensitivity of the removed streams, but in order to keep computational complexity down, we did not explore this option in this paper. 

\begin{figure}
  \begin{center}
  $\vcenter{\hbox{\includegraphics[scale=0.25]{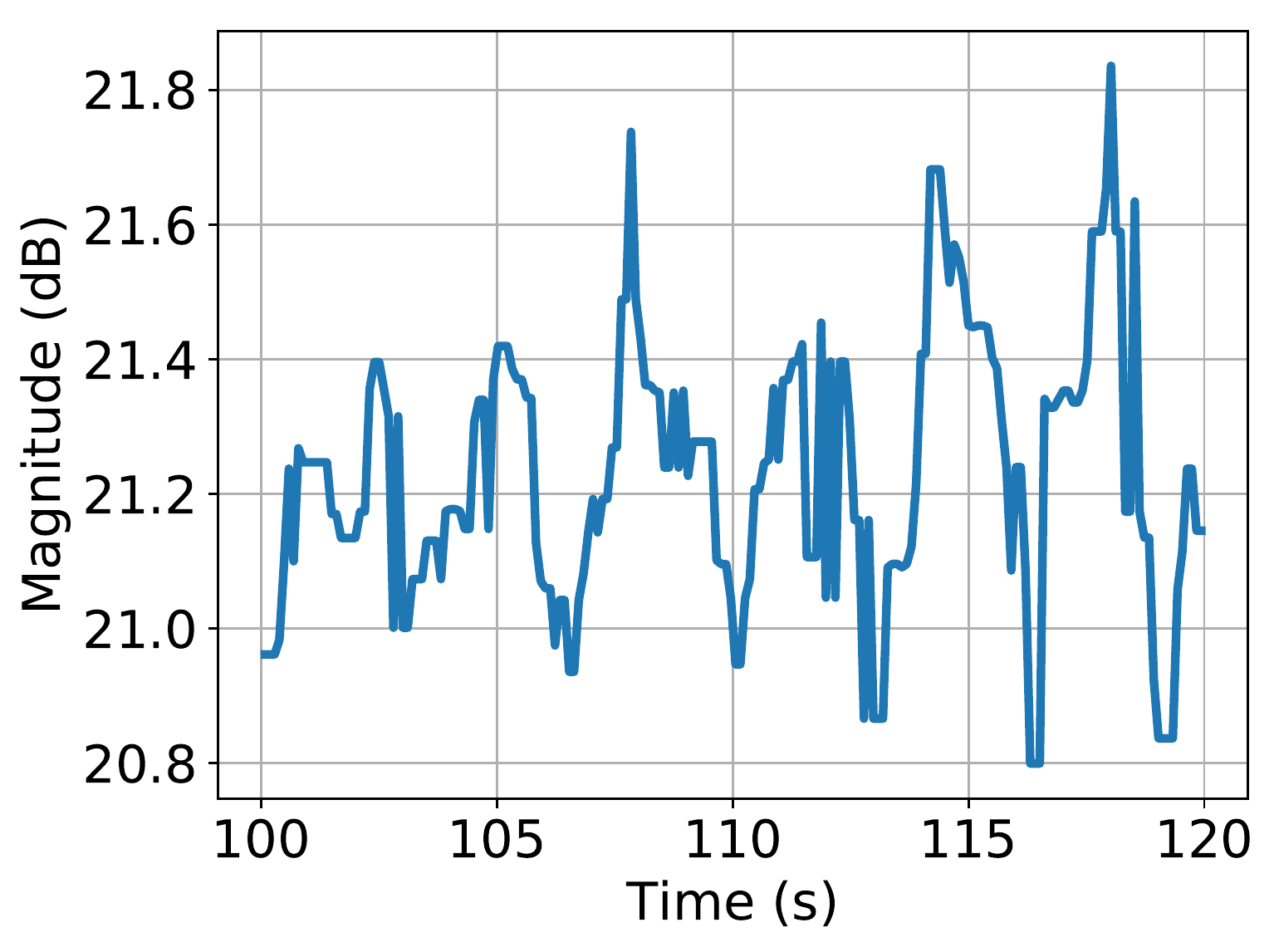}}}$\quad
  $\vcenter{\hbox{\includegraphics[scale=0.25]{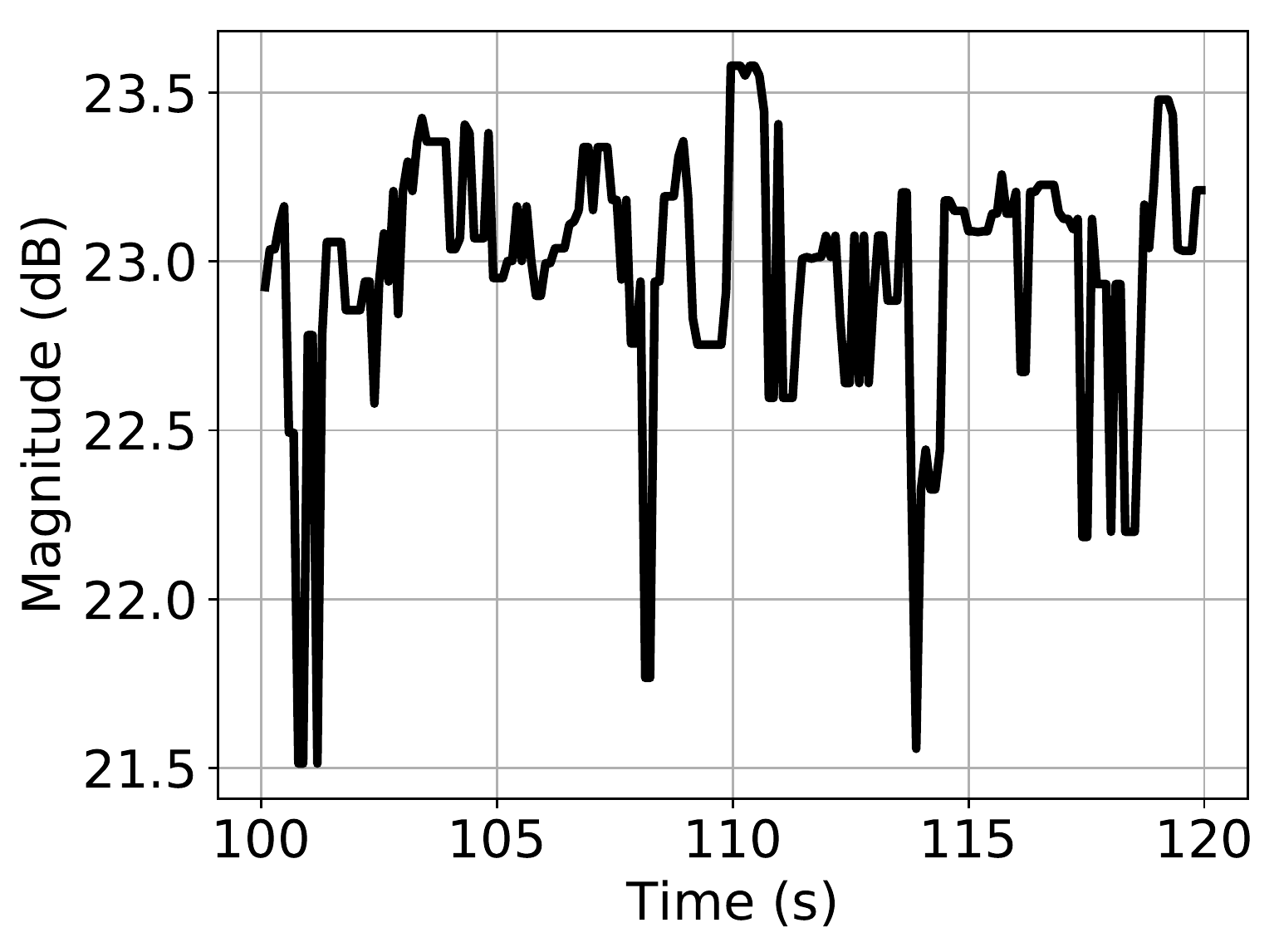}}}$
  \caption{Subcarriers plotted over the same time period when a patient was breathing at 18 bpm. (Left) A subcarrier from the subcarrier group with low variance. (Right) A subcarrier from the subcarrier group with high variance.}
  \label{F:noisy_clean_subcarriers}
  \end{center}
  \vspace{-1em}
\end{figure}

{\bf Channel State Information}:
While developing algorithms for CSI, we observed that subcarriers from one receiver antenna tended to have a higher variance than the subcarriers from the other receiver antenna. We show $\mathbf{y}^{csi}$ for two subcarriers, one from each receiver antenna, in Fig.~\ref{F:noisy_clean_subcarriers}.
 The group with the higher variance can change after a person moves. We found that removing the subcarriers in the group with the higher variance improves estimation and detection. We acknowledge that this is counterintuitive since higher signal variance would suggest a clearer respiration signal \cite{baboli2012respiration}. Though we do not know the fundamental reasoning, we attribute the higher variance to higher noise variation. The process of filtering out the high-variance subcarrier group is done by first grouping subcarriers by receiver and computing the moving variance of each subcarrier over a 30~\si{\second} window. We then compute the minimum, mean, median, and maximum variance for each subcarrier group. A subcarrier group gains a point if the minimum, mean, median, and maximum variance is less than the minimum, mean, median, and maximum variance of the other group. The subcarrier group with the least number of points are filtered out. In case of a tie, we filter out the same subcarrier group that was filtered out during the previous sample. 

{\bf 1-dB Quantized RSS}:
Some streams of 1-dB quantized RSS measurements can have very high noise that overwhelms any possible respiration signal. These streams have very high variances. In contrast, the changes in RSS due to respiration can fall completely within a 1-dB quantization level. As a result, these streams contain no respiration signal. To filter out these two groups, we compute the variance of the RSS for each stream over a window and only keep the links with a variance between the $25^{th}$ and $75^{th}$ variance percentile. 

{\bf Sub-dB Quantized RSS}:
At the time of this paper, we only measured sub-dB quantized RSS on one channel and therefore do not need to use stream selection. However, we hypothesize that applying a similar stream selection method like those presented for CIR, CSI, and RSS would be useful had sub-dB RSS on more channels been measured. 

%
%
\subsection{Signal Filtering} \label{S:rf_generic_sp}
The initial processing performed for each RF technology yields a measurement vector $\mathbf{y}$. A person's chest moving during inhalation and exhalation cause sinusoidal changes to $\mathbf{y}$. We filter these measurements to remove unwanted high and low frequency components in the measurements.

On average, the respiratory rate of a healthy adult at rest is 14 bpm \cite{sebel1985respiration}, and can vary from 12-15 bpm \cite{barrett2012ganongs}. We consider that higher frequency components in the signal are caused by motion other than respiration or noise which we desire to filter out. We create a fifth-order Butterworth low-pass filter with a cutoff frequency of 0.4~\si{\hertz} to attenuate high frequency signals. The Butterworth filter's flat frequency response in the passband is desirable since it does not amplify any specific frequencies in the passband.

\begin{figure}
  \begin{center}
  $\vcenter{\hbox{\includegraphics[scale=0.25]{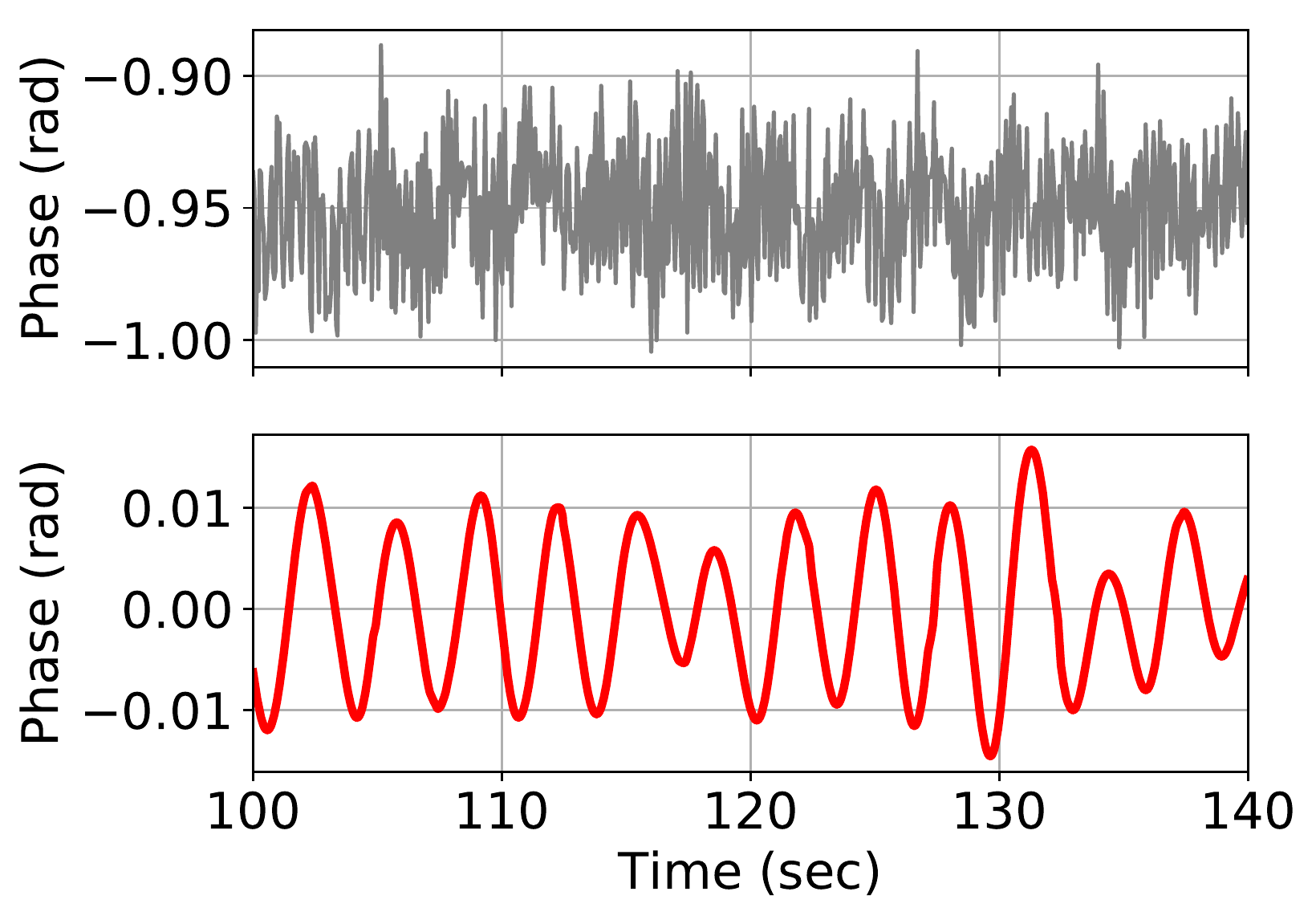}}}$
  $\vcenter{\hbox{\includegraphics[scale=0.25]{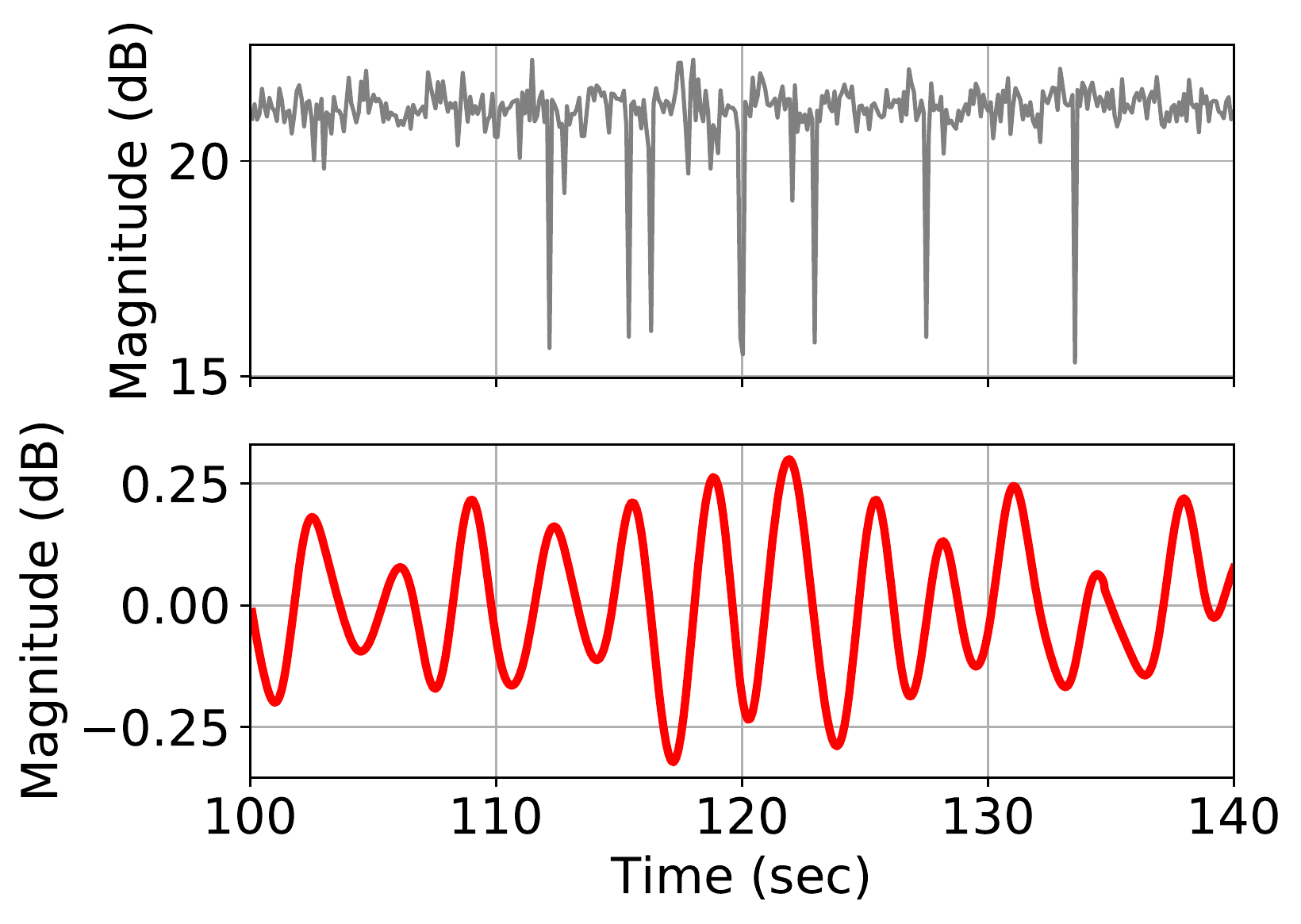}}}$
  $\vcenter{\hbox{\includegraphics[scale=0.25]{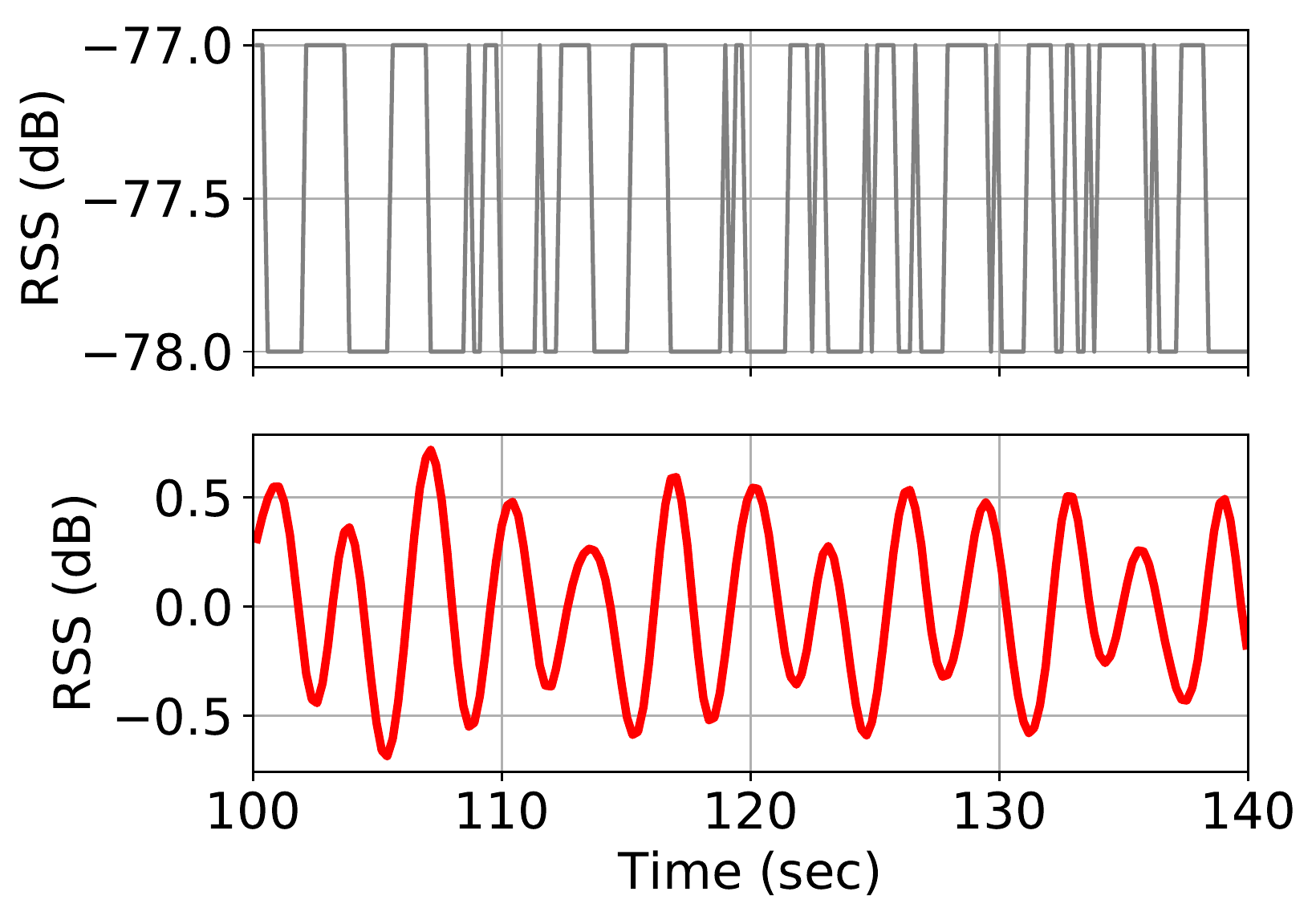}}}$
  $\vcenter{\hbox{\includegraphics[scale=0.25]{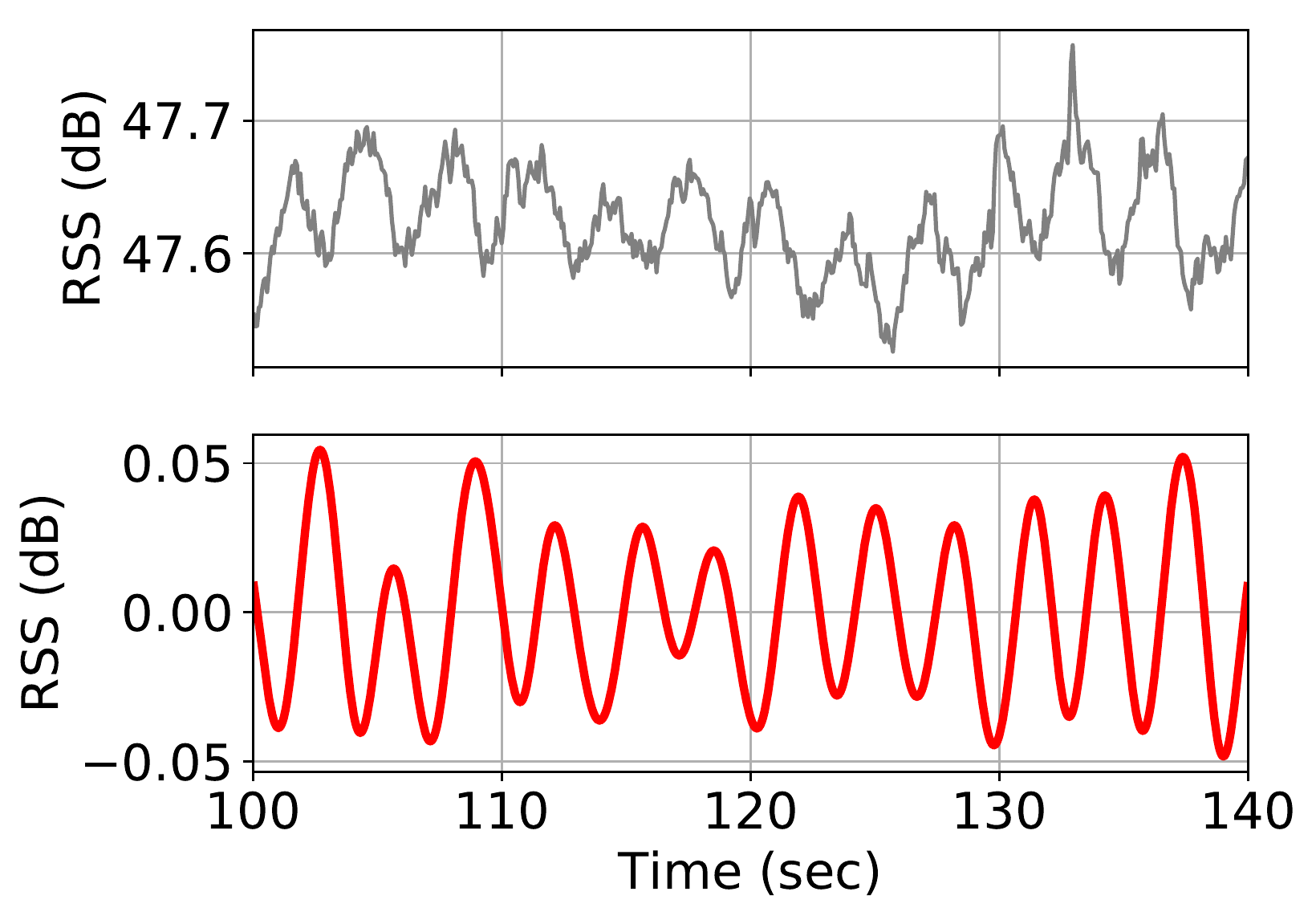}}}$
  \caption{A 1 min interval of the raw  $\mathbf{x}$ in gray and filtered signal $\tilde{\mathbf{y}}$ in red for (top-left) CIR (top-right) CSI (bottom-left) 1-dB quantized RSS (bottom-right) sub-dB quantized RSS measurements for a person breathing at 18 breaths/min.}
  \label{F:filtered_timeseries}
  \end{center}
  \vspace{-1em}
\end{figure}

After running $\mathbf{y}$ through a low-pass filter, we run the measurements through a high-pass filter to obtain a zero-mean signal. This is a necessary step when computing the power spectral density (PSD) of a noisy, finite length signal since the DC component can overwhelm the power of lower amplitude sinusoidal components. We discuss the PSD in Section \ref{S:respiration_estimation}. The high-pass filter is a fifth-order Butterworth filter with a cutoff frequency of 0.1~\si{\hertz}. We denote the measurement after the low and high-pass filter as $\tilde{\mathbf{y}}$. The result of filtering for all RF measurements are shown in Fig.~\ref{F:filtered_timeseries}.

%
%
\subsection{Respiratory Rate Estimation} \label{S:respiration_estimation}
As seen in Fig.~\ref{F:filtered_timeseries}, the processed and filtered RF measurements have a sinusoidal component when the person is breathing. There are two popular ways of estimating the RR from these signals. In the first method, the average PSD is computed using a 10 - 30~s window of measurements \cite{patwari2014breathfinding}. The frequency at which the PSD is maximum is the estimated RR. We denote this estimated RR as $\hat{f}_{psd}$~\si{\hertz}. In this paper, we use a 30~s window of measurements and compute the PSD between $f_{min}=0.1$~\si{\hertz} and $f_{max}=0.4$~\si{\hertz} with a step size of $f_{step}=0.002$~\si{\hertz}. $f_{min}$ and $f_{max}$ are set to account for a range of breathing rates for a resting healthy adult.

The second method is to compute the inverse of the time between inhalations \cite{liu2015vital}, known as the inter-breath interval (IBI). In this method, a peak finding algorithm finds the peaks of $\tilde{\mathbf{y}}$ over a 10 - 30~\si{\second} window. The mean time difference between the peaks for each stream is computed. The inverse of the average of the means is then computed to get the estimated RR. We denote this estimated respiration rate as $\hat{f}_{ibi}$~\si{\hertz}. In this paper, we implement the IBI algorithm presented in \cite{liu2015vital} which applies peak filtering and an additional weighting method to compute $\hat{f}_{ibi}$.

For both the PSD and IBI methods, a new $\hat{f}$ estimate is produced every ~\SI{5}{\second}. We compare the accuracy of the PSD and IBI methods later in Section \ref{S:psd_vs_ibi}.

\subsection{Ground Truth Respiration Rate}
In each polysomnography study, the patient is monitored using a variety of sensors including respiratory inductance plethysmography (RIP) belts around the chest and abdomen, and a thermocouple and nasal pressure sensor placed in the nose. The RIP belts measure the chest and abdomen expanding and contracting during breathing. The thermocouple measures changes in the temperature related to inspiration and exhalation while the nasal pressure sensor measures changes in pressure related to the same.

\begin{figure}
  \begin{center}
  $\vcenter{\hbox{\includegraphics[scale=0.245]{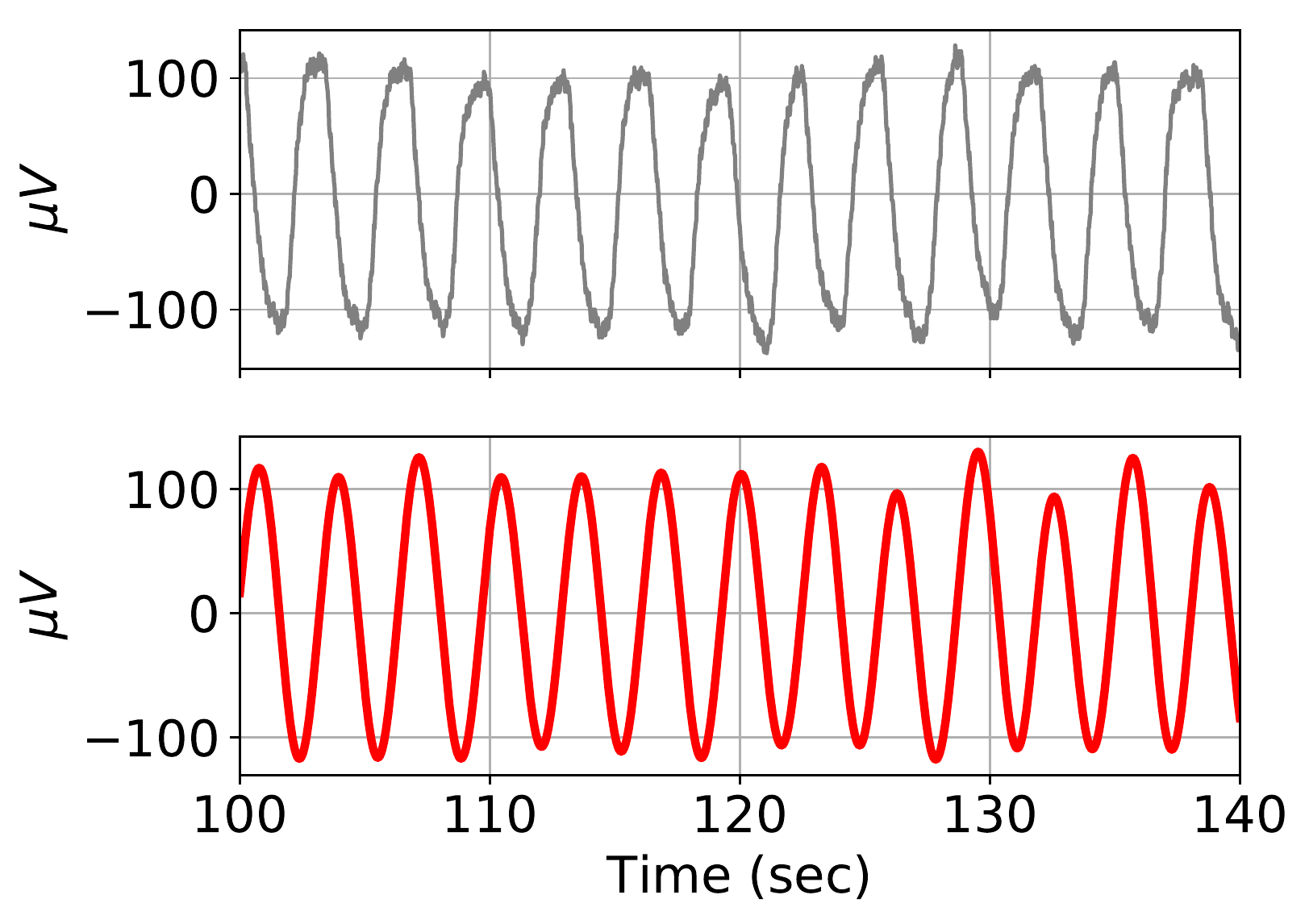}}}$
  $\vcenter{\hbox{\includegraphics[scale=0.245]{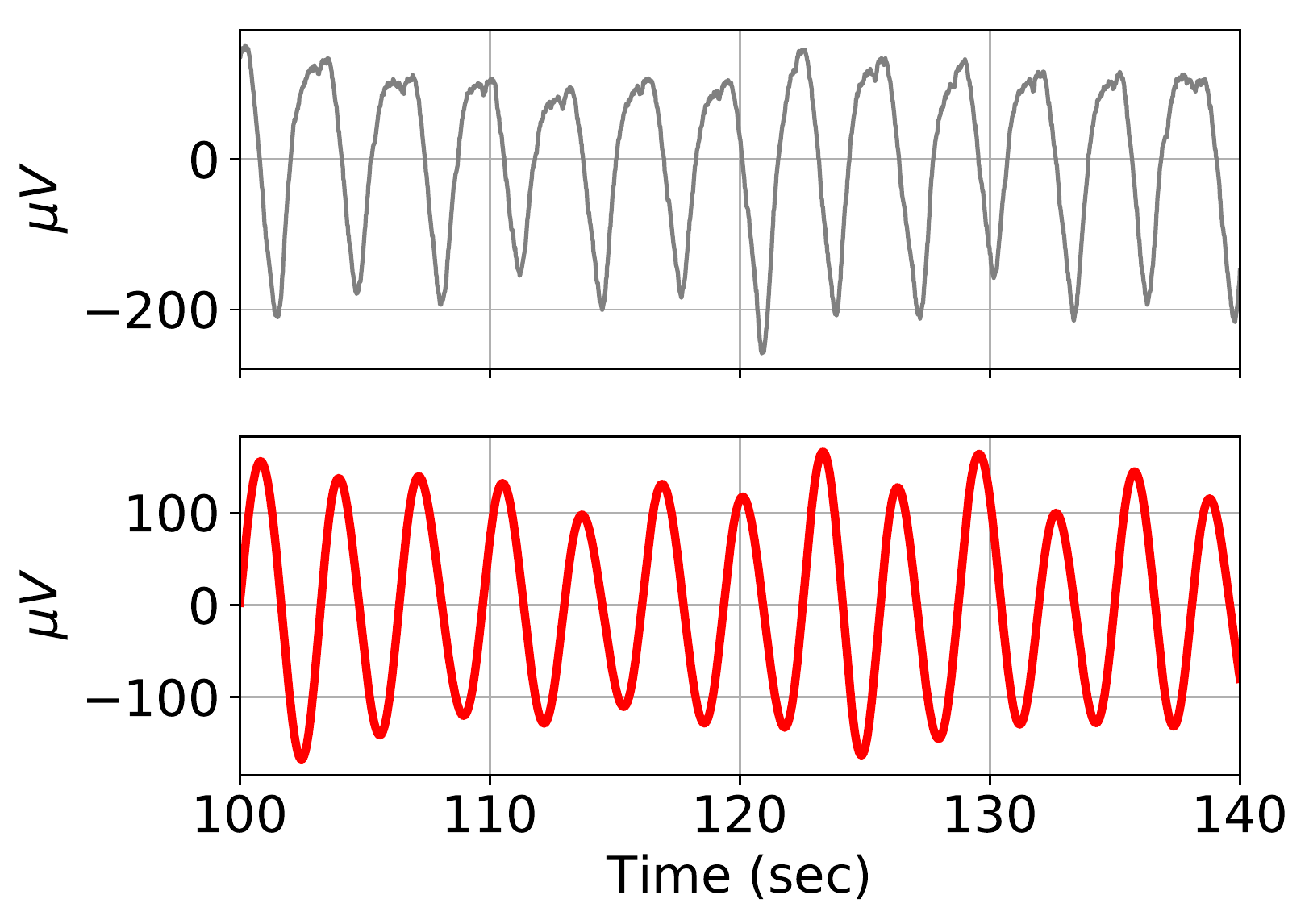}}}$
  $\vcenter{\hbox{\includegraphics[scale=0.245]{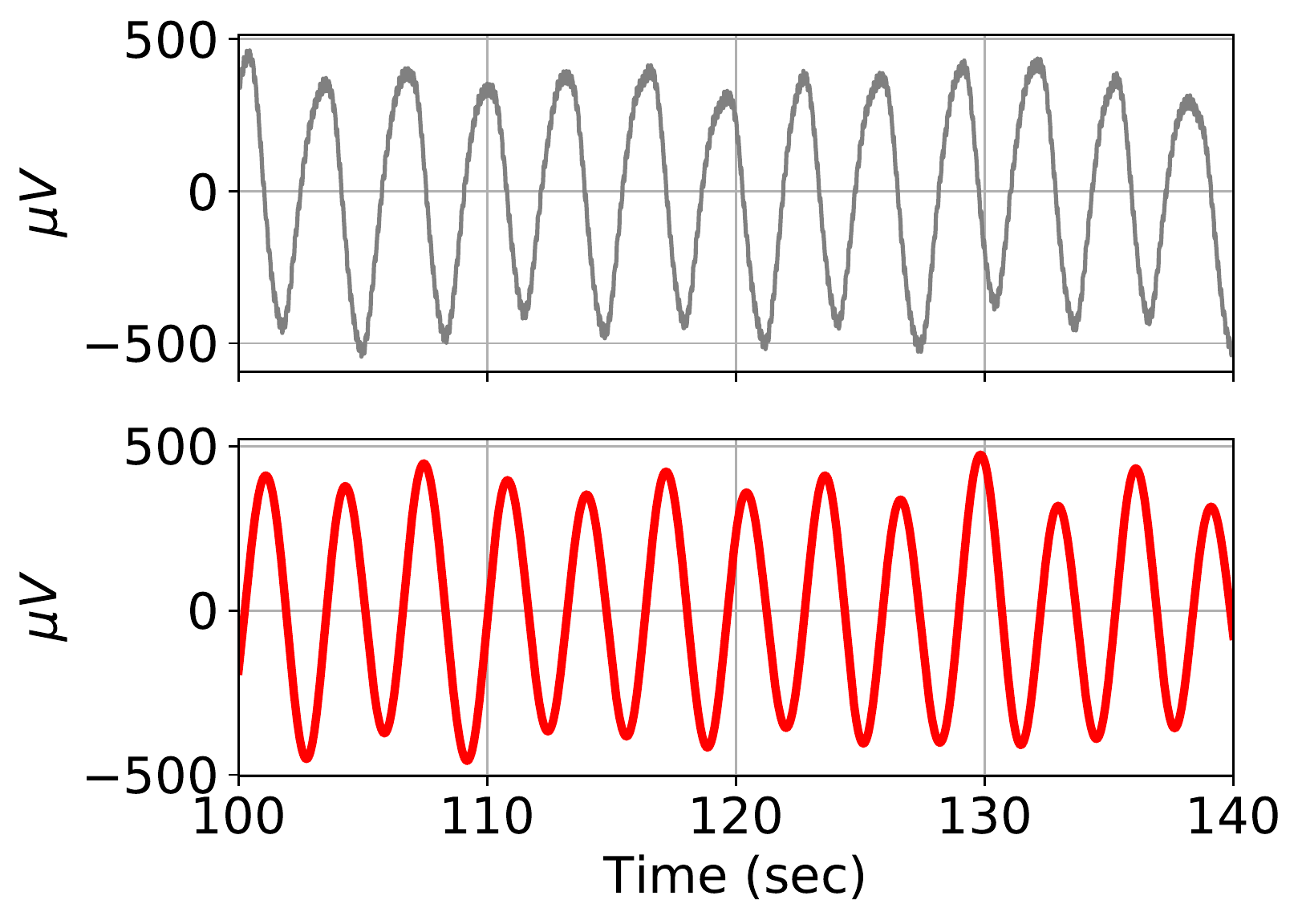}}}$
  $\vcenter{\hbox{\includegraphics[scale=0.245]{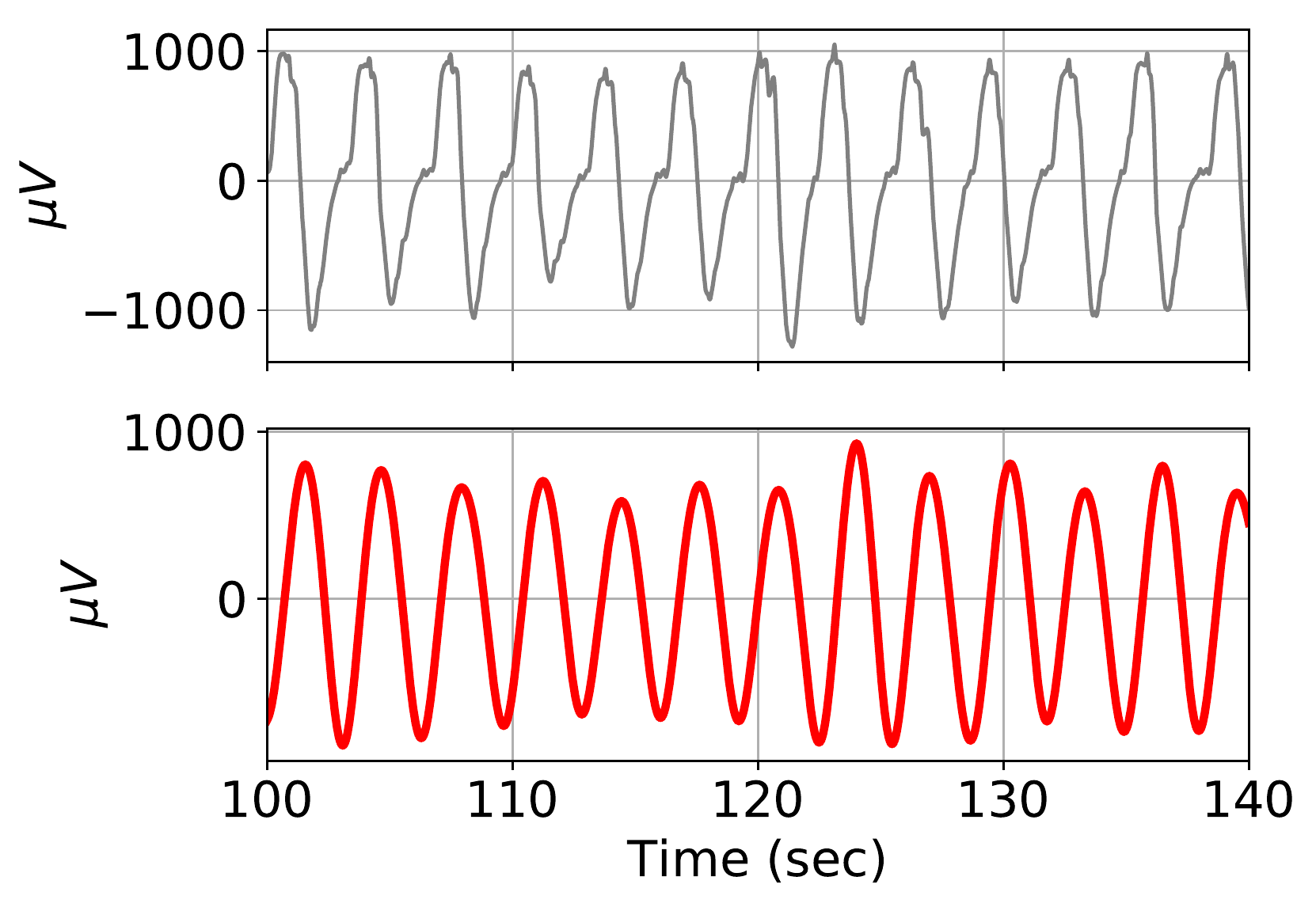}}}$
  \caption{A 1 min interval, for a person breathing at 18 breaths/min of the raw  $\mathbf{x}^{poly}$ in gray and filtered signal $\tilde{\mathbf{y}}$ in red for (top-left) chest RIP band (top-right) abdomen RIP band (bottom-left) thermocouple (bottom-right) nasal pressure measurements, used to establish the RR ground truth.}
  \label{F:poly_msrts}
  \end{center}
  \vspace{-1em}
\end{figure}

One measurement from each of these four sensors form the measurement vector $\mathbf{y}^{poly} \in \mathbb{R}^{4}$ and are sampled at $f_s^{poly} = 25$~\si{\hertz}. The measurements are then sent through a low-pass and high-pass filter as was described in Section~\ref{S:rf_generic_sp} to form the vector $\tilde{\mathbf{y}}^{poly}$. A time series of these four sensors and their filtered form are shown in Fig.~\ref{F:poly_msrts}. 
 The ground-truth respiration rate is then estimated using the PSD estimation solution described in Section~\ref{S:respiration_estimation}.

\subsection{Motion Detection} \label{S:motion_detection}
During respiration monitoring, a patient may roll to a new sleeping position or may some part of their body. The changes in the RF channel measurements during these periods can overwhelm the small changes from their respiration. As a result, it is very difficult to estimate the patient's RR when the signal is dominated by motion. Motion detectors provide a way to flag periods of time when the RR estimate is not reliable. In this section, we describe five different motion detection methods. The first method, which we refer to as a moving average-based detector (MABD), computes the absolute percent change between a long-term and short-term average as described in \cite{youssef2007challenges}. A second is a moving variance-based detector (MVBD) which computes the absolute difference between a short-term and long-term variance \cite{youssef2007challenges}. A third detector computes the average variance energy (AVE) over a short window \cite{liu2015vital}. For these three detectors, motion is detected when the value exceeds a threshold. A fourth detector computes the ratio of the maximum amplitude to the average amplitude of the frequency response \cite{adib2015breathing}. The closer the ratio is to one, the less periodic the channel measurements are during the most recent window of measurements. Motion is detected when the ratio falls below a threshold. We refer to this detector as the flat spectrum detector (FSD). The fifth detector combines the MABD and MVBD by detecting motion when both MABD and MVBD detect motion. We refer to this fifth detector as MAVBD. The purpose of this detector is to reduce the number of false alarms through majority vote.

In this paper, we evaluate the usefulness of motion detection algorithms for long-term respiration rate monitoring by comparing performance when motion detection is and is not used. When motion detection is used, we also compare how each method performs against the other. When motion is detected, we ignore the respiration rate estimate by letting $\hat{f}=NaN$. We compare these different settings in Section~\ref{S:results}.

%% file: results.tex
\section{Results} \label{S:results}
In this section, we compare the RR estimation as a function of methods used in the blocks shown in Fig.~\ref{F:breathing_monitor_system}. We apply different methods in each of the blocks and then make comparisons based on metrics defined in the sections that follow.

\subsection{Performance Metrics}
The performance of each RF technology and the methods used are compared using the absolute difference between the true and estimated respriation rate. This metric is formalized as
\begin{equation}
    e = 60\cdot|f_{gt} - \hat{f} |
\end{equation} where $f_{gt}$ is the ground truth frequency in~\si{\hertz} and $e$ is in bpm. The polysomnography and each RF technology have a different sampling frequency, and so $e$ is computed using the $f_{gt}$ and $\hat{f}$ measured at the closest points in time. When a motion detection algorithm is used, as described in Section~\ref{S:motion_detection}, $\hat{f} = NaN$ and so we do not compute $e$.



\subsection{Stream Selection's Effect on RR Estimation}
\begin{figure}[tb]
  \begin{center}
  \includegraphics[width=0.90\columnwidth]{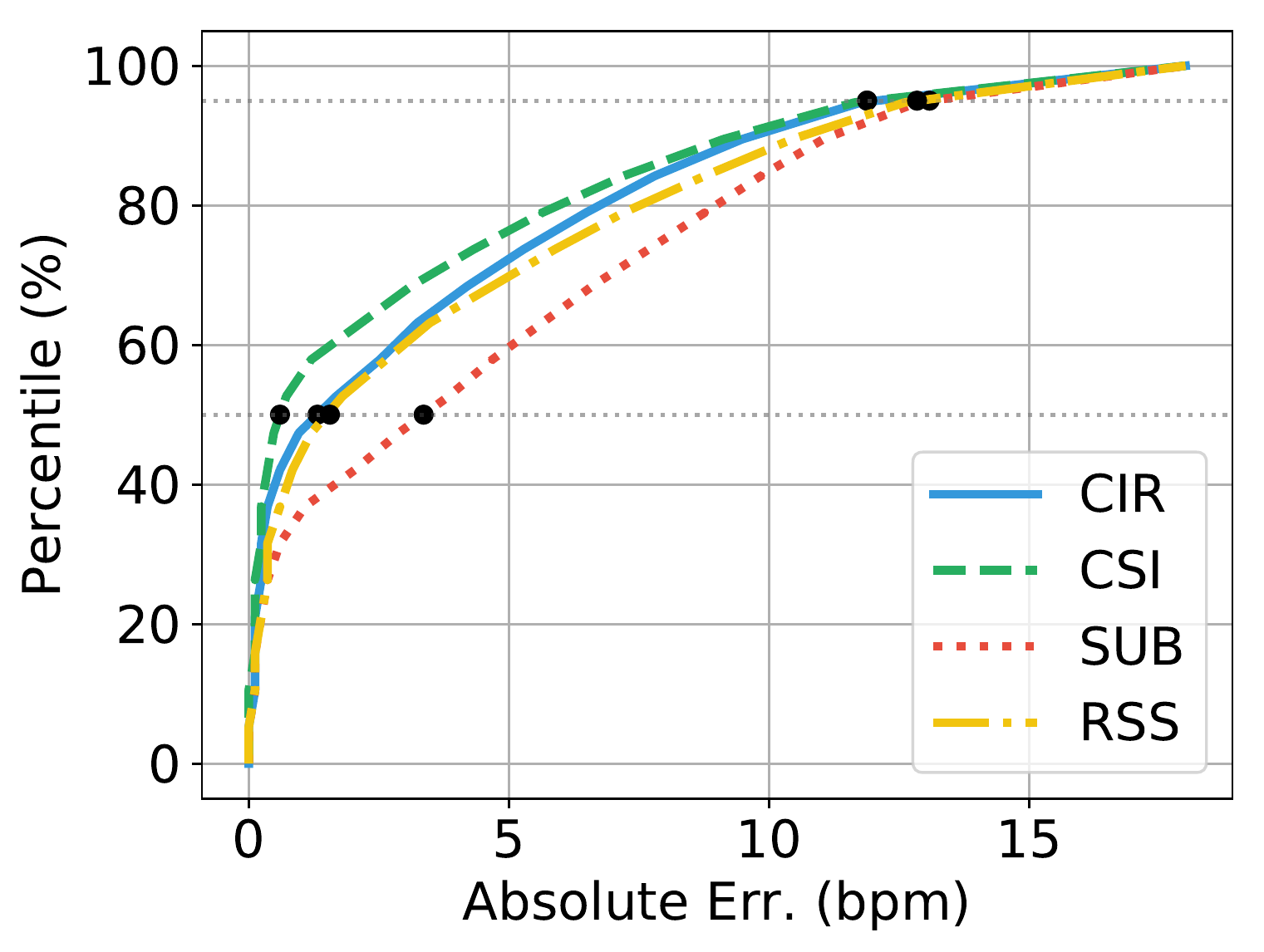}
  \caption{CDF of $e$ over all 20 studies when stream select is enabled. A PSD method is used but motion detection is bypassed.}
  \label{F:err_stream_select_PSD}
  \end{center}
  \vspace{-1em}
\end{figure}

\begin{figure}[tb]
  \begin{center}
  \includegraphics[width=0.90\columnwidth]{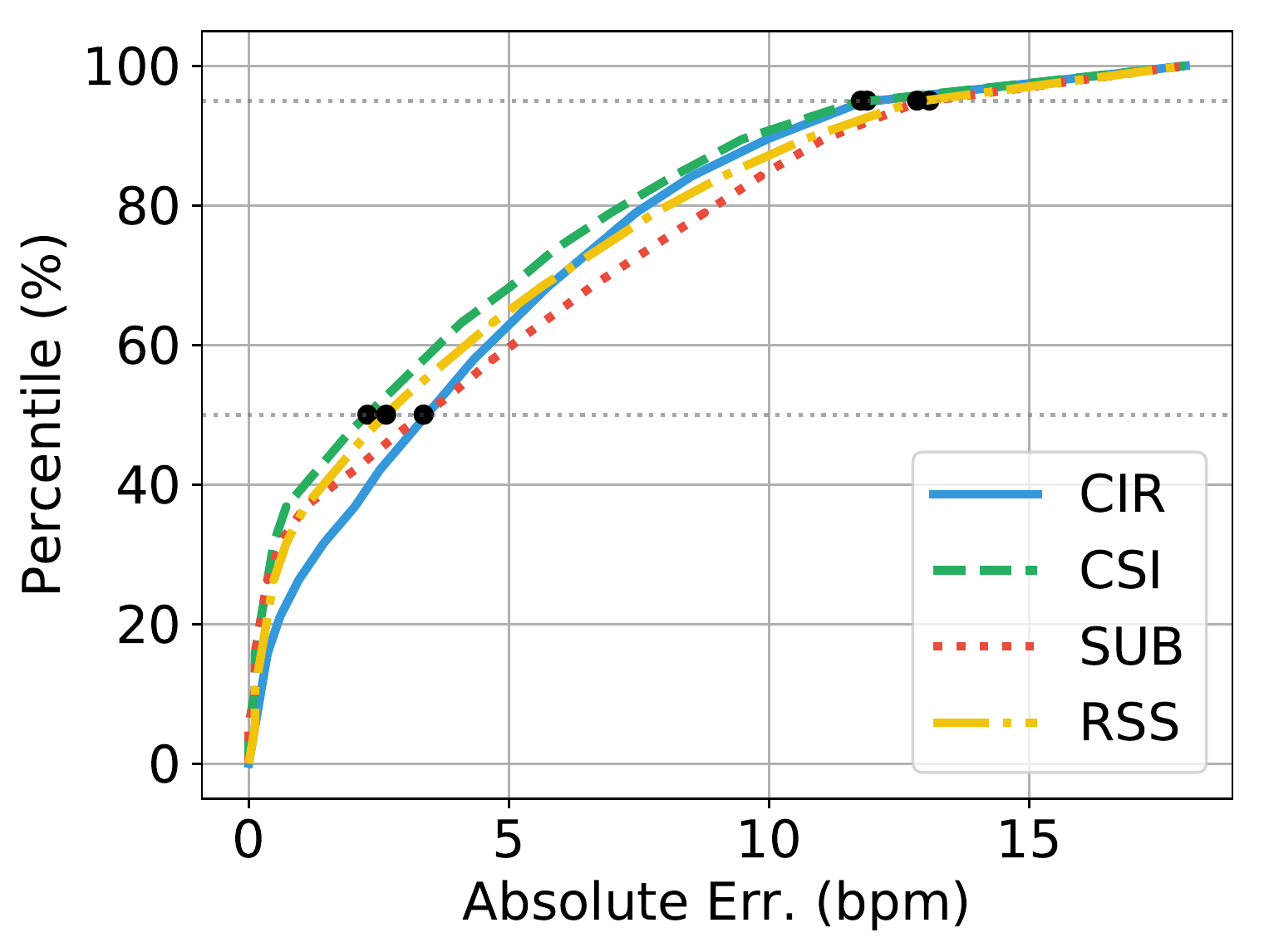}
  \caption{CDF of $e$ over all 20 studies when stream select is bypassed. A PSD RR method is used but motion detection is bypassed.}
  \label{F:err_stream_select_bypass_PSD}
  \end{center}
  \vspace{-1em}
\end{figure}

Selecting the best streams or weighting streams based on their sensitivity to breathing has been applied in some research, but not others. In this section, we compare the RR estimation error when stream selection is enabled and bypassed to show the effect of stream selection on the RR estimation error. We use the PSD RR estimation method and we disable the motion detection block. In Figs.~\ref{F:err_stream_select_PSD} and \ref{F:err_stream_select_bypass_PSD}, we observe the CDF of $e$ over all twenty studies.
For CIR, CSI, and RSS, the CDF when stream selection is enabled shows a lower $e$ for a greater percentage of the time when compared to when stream selection is bypassed. There is no difference between the CDFs for SUB since only one channel was measured. These differences between the errors are more apparent in Table~\ref{T:stream_SEL} which shows the median $e$ of each RF technology and for each stream selection mode.
\begin{table}
\centering
\caption{The median $e$ for each RF technology when stream selection is enabled and bypassed. The Mann-Whitney U p-value is provided.}
\label{T:stream_SEL}
\begin{tabular}{c|c|c|c}
 & \multicolumn{2}{c|}{Stream Selection} & \multicolumn{1}{c}{Mann-Whitney U} \\
RF Tech & \multicolumn{1}{c|}{Enabled} & \multicolumn{1}{c|}{Bypassed} & p-value \\ \hline \hline
CIR & 1.32 & 3.36 & $<$~\num{1e-6} \\ 
CSI & 0.60 & 2.28 & $<$~\num{1e-6} \\  
RSS & 1.56 & 2.64 & $<$~\num{1e-6} \\
SUB & 3.36 & 3.36 & N/A \\
\end{tabular}
\vspace{-1em}
\end{table} 

From the CDFs and the table, we see significant improvement in respiratory rate estimation when stream selection is enabled. The median $e$ drops by 2 bpm for CSI and CIR and 1 bpm for RSS. To check if a randomly selected $e$ from the enabled stream select mode is less than a randomly selected $e$ from the bypassed stream select mode, we run a Mann-Whitney U test. In this test, the null hypothesis is the $e$ from both stream select modes are drawn from the same distribution. The alternative hypothesis is that the distributions are not equal. For CIR, CSI, and RSS, the p-values are all $<$~\num{1e-6}. The results of these tests suggest that there is statistical significant difference between the $e$ from the stream select enabled and bypassed mode. Selecting the best streams is an important processing step that can lead to significant improvement in RR estimation.

Another important observation we make is that there is a difference between CDFs when we compare each RF technology with stream selection enabled. CSI achieves the lowest median $e$, two to five times lower than the median $e$ for CIR, RSS, and SUB. Stated differently, when a person is breathing at $\geq 12$ bpm, the median $e$ when using CSI will be within $\geq 95\%$ of the true RR. Why CSI achieves a lower median $e$ compared to the other wireless devices, and CIR in particular which uses the most bandwidth, could be a function of its frequency diversity, making use of antenna diversity by measuring a MIMO instead of SISO channel, the larger number of streams used for RR estimation, and/or its coherence bandwidth. These and other possibilities could be investigated to provide a more theoretical reasoning for CSI performing the best out of the four RF devices.

However, these results don't tell the full story. While CSI achieves the lowest median $e$, the CDFs show that all four RF devices are all able to achieve very low RR estimation errors. For CSI, 56\% of the estimates are less than 1 bpm, 47\% for CIR, 44\% for RSS, and 36\% for SUB. All technologies can achieve low RR errors, but they all vary in how often the RR estimate is accurate. The other side to this story is that there is a significant number of RR estimates which are unreliable for all four RF devices. In the following sections, we continue to evaluate how different methods affect RR estimation on a processing block basis. But we also offer observations of why the RR estimate can be so unreliable.

\subsection{Estimation Method's Effect on RR Estimation Accuracy} \label{S:psd_vs_ibi}
\begin{figure}
  \begin{center}
  \includegraphics[width=0.90\columnwidth]{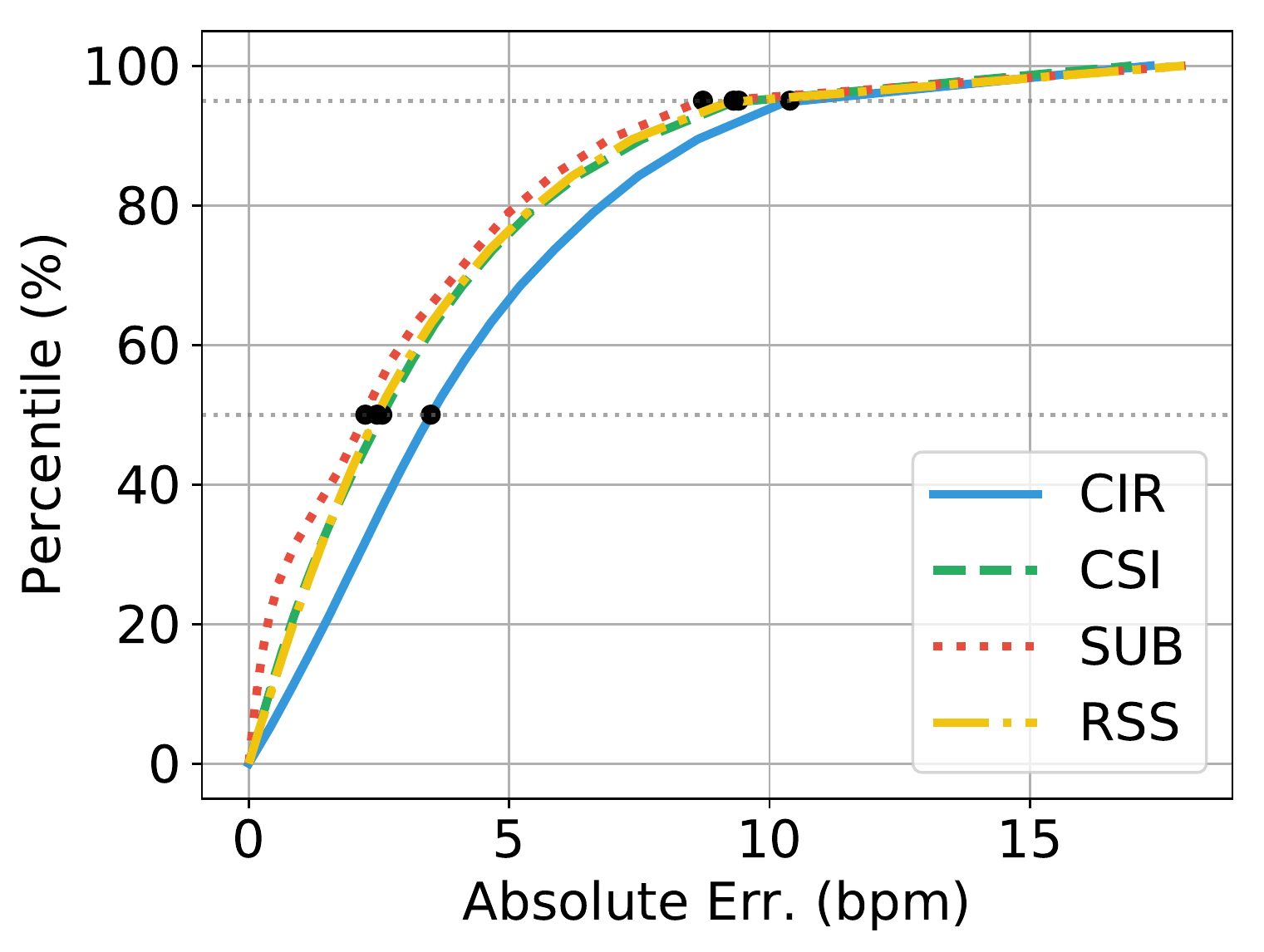}
  \caption{CDF of $e$ over all 20 sleep studies when the IBI RR estimation method is used.}
  \label{F:err_vIBI}
  \end{center}
  \vspace{-1em}
\end{figure}

As explained in Section~\ref{S:respiration_estimation}, two popular RR estimation methods are computing the peak frequency from an average PSD or computing the IBI from the time domain signals. In this section, we compare the PSD approach as explained in \cite{patwari2014breathfinding,patwari2014breathing,kaltiokallio2014respiration} and the IBI method explained in \cite{liu2015vital}. In these evaluations, stream selection is enabled, but no motion detection method was used. In Fig.~\ref{F:err_vIBI}, the CDFs for each RF technology is shown when the IBI respiration rate estimation method is used.
 A summary of the CDFs in Fig.~\ref{F:err_vIBI} and in Fig.~\ref{F:err_stream_select_PSD} are shown in Table~\ref{T:RR_est_method} which shows the median and 95th percentile of $e$ for the PSD and IBI methods.
\begin{table}
\centering
\caption{The median / 95th percentile $e$ for each RF technology when stream selection is enabled and bypassed.}
\label{T:RR_est_method}
\begin{tabular}{c|c|c}
RF Tech & PSD & IBI \\ \hline \hline
CIR & 1.32~/~11.88 & 3.49~/~10.39 \\ 
CSI & 0.60~/~11.88 & 2.57~/~9.41 \\  
RSS & 1.56~/~12.84 & 2.46~/~9.31 \\
SUB & 3.36~/~13.08 & 2.24~/~8.72 \\
\end{tabular}
\vspace{-1em}
\end{table} When we visually compare the CDFs in Fig.~\ref{F:err_vIBI} to the CDFs in Fig.~\ref{F:err_stream_select_PSD}, we observe the accuracy for CIR, CSI, and RSS degrades when IBI is used and when $e<5$ bpm. Specifically, the percent of RR estimates that are $<1$ bpm is 13\% for CIR, 24\% for CSI, and 23\% for RSS. The same statistic for SUB is 32\%. While IBI may not be the RR estimation method for RSS, CIR, and CSI, there is improvement for SUB when using IBI instead of the PSD method. The median error decreases by 0.9 bpm. However, SUB does not improve its percent of estimates $<1$ bpm.

Why does IBI perform better with SUB but not the other RF devices? Note in this scenario, better is purely relative, because the median $e$ for SUB still is impractically high to be useful in most applications. Furthermore, the percent of estimates $<1$ bpm for SUB does not improve when using IBI. But to address the question, we note that the $95th$ percentile for all RF devices is $1.5$ to $4.0$ bpm lower when IBI is used than when PSD is used. The IBI algorithm may have been designed in a way such that large $e$ were avoided whereas the same cannot be said of the PSD method. The way the algorithm operates could naturally make it appear that IBI works better for SUB. But, since the results are still poor, we conclude that the PSD method is a more robust RR estimation method. 

\subsection{Motion Detection Effect on RR Estimates}
\begin{figure}
  \begin{center}
  $\vcenter{\hbox{\includegraphics[scale=0.25]{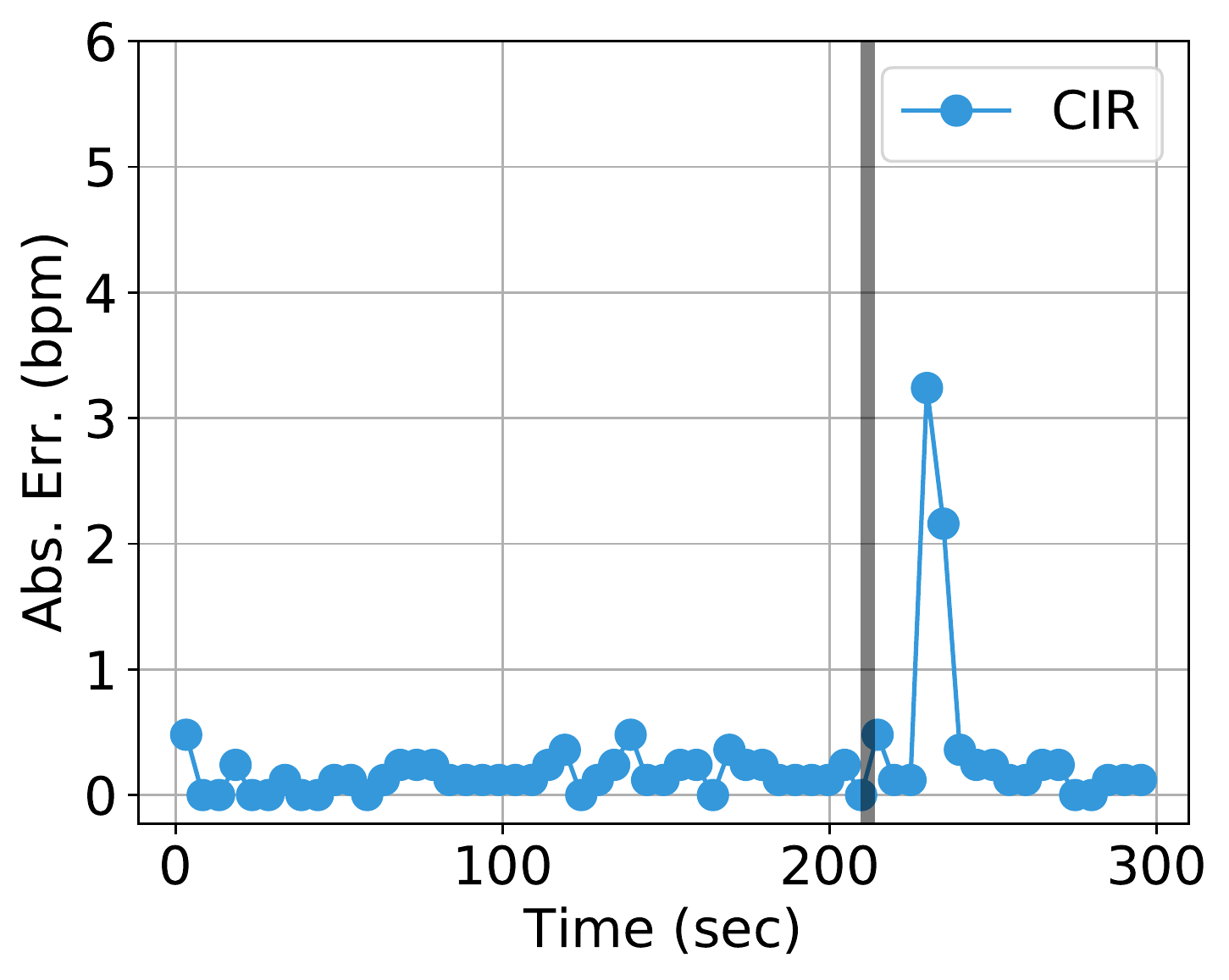}}}$\quad
  $\vcenter{\hbox{\includegraphics[scale=0.25]{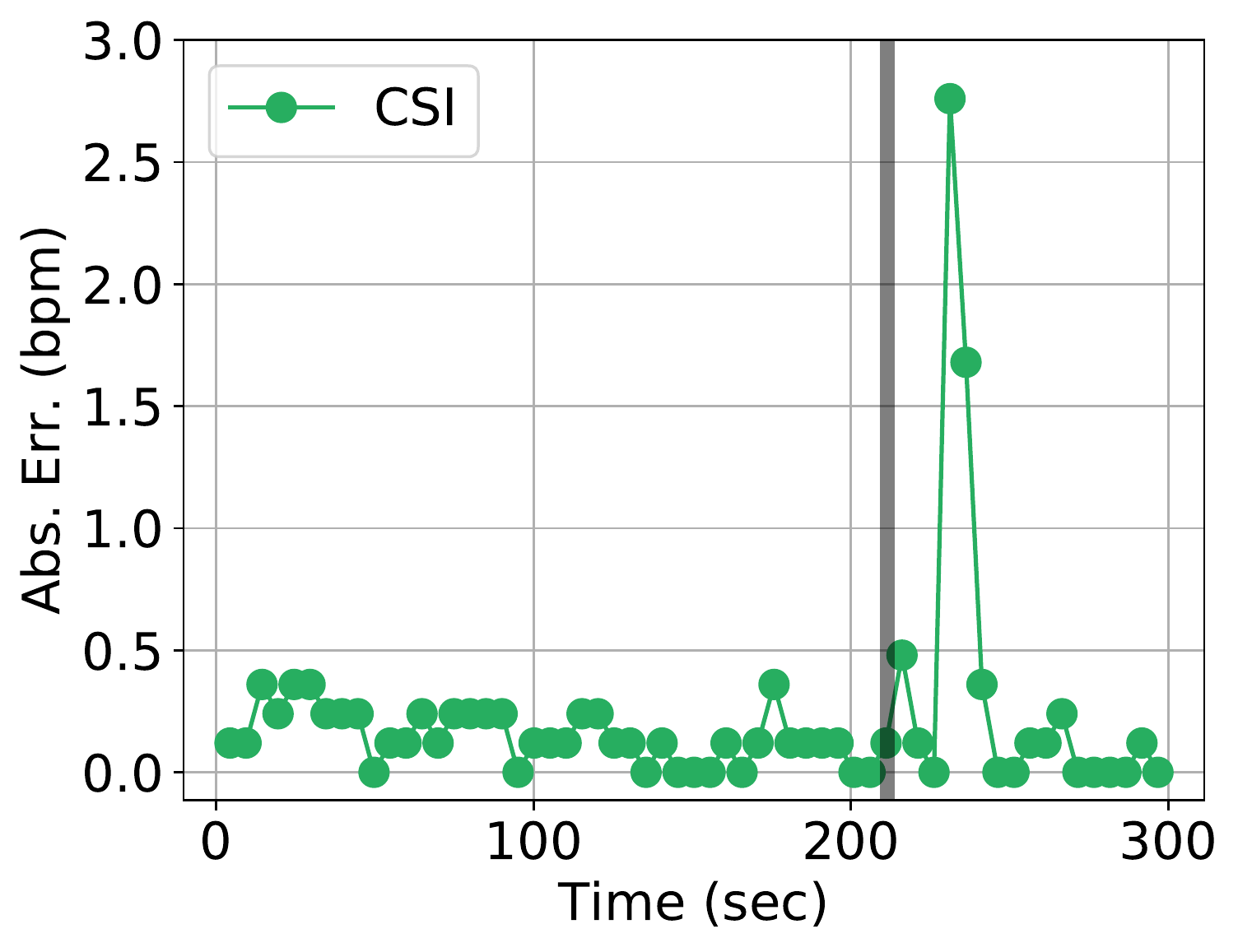}}}$\quad
  $\vcenter{\hbox{\includegraphics[scale=0.25]{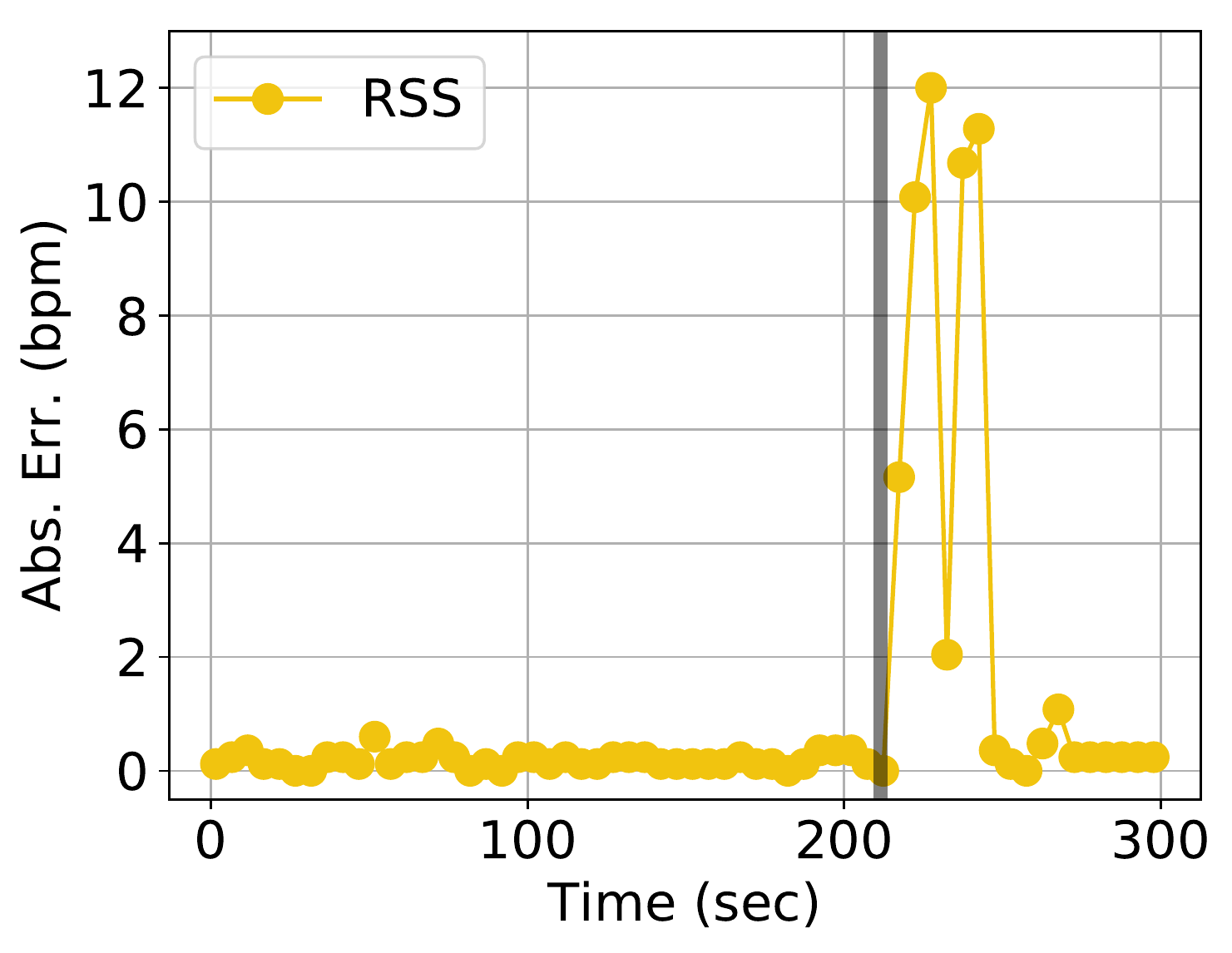}}}$\quad
  $\vcenter{\hbox{\includegraphics[scale=0.25]{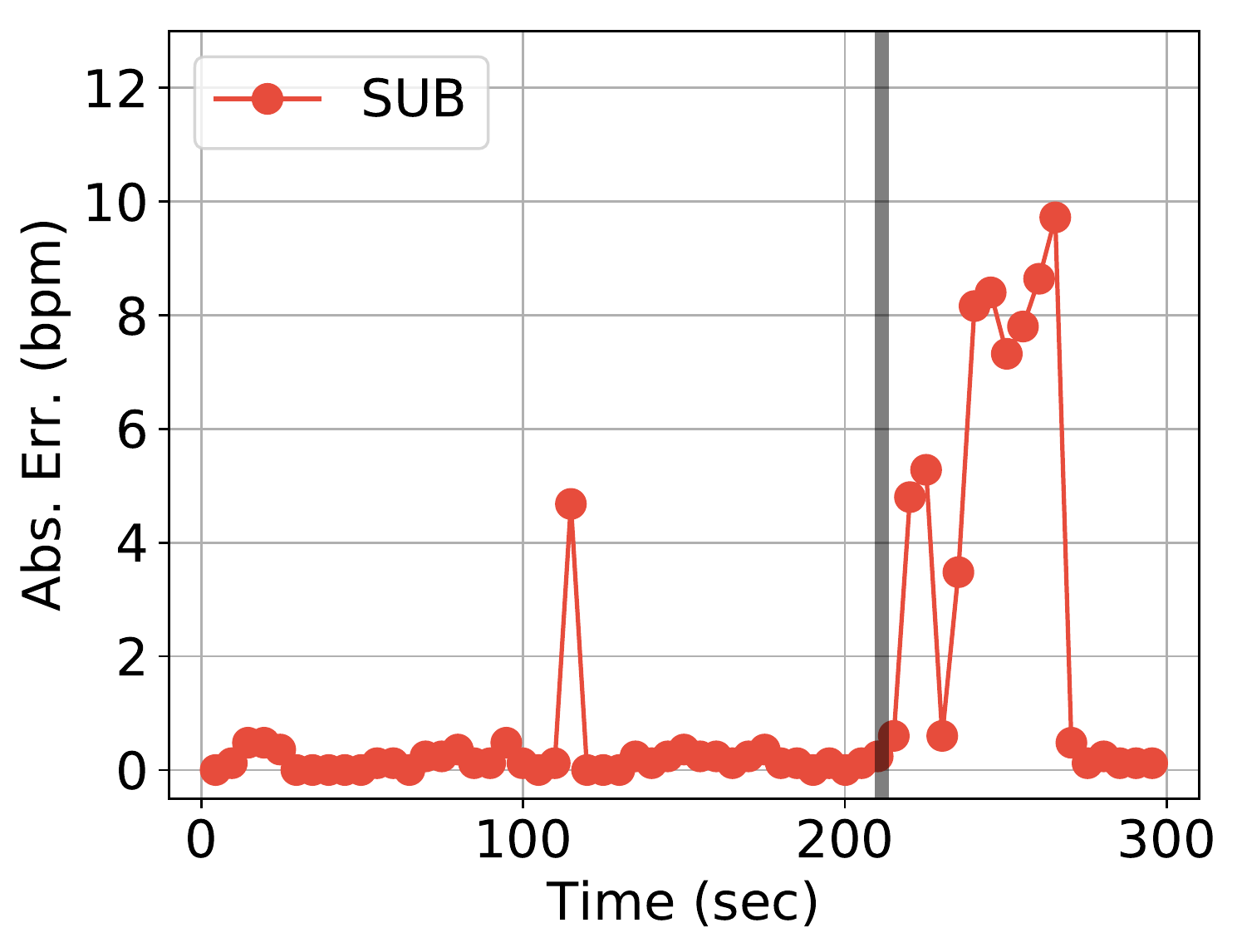}}}$
  \caption{A five minute interval showing the absolute error achieved by each technology when no motion detection algorithm is used. The gray band shows a period when motion occurred.}
  \label{F:motion_err_no_motion_detection}
  \end{center}
  \vspace{-1em}
\end{figure}

So far, we have not included the output of a motion detection method while estimating respiration rate. Motion periods can result in large changes in the channel measurements which overwhelm the small changes caused by respiration. Motion events can, in turn, cause large errors in the respiratory rate estimation. This can be seen in Fig.~\ref{F:motion_err_no_motion_detection}.
 In this figure, we observe that there are a few respiration rate estimate errors above 1.5. The number of high error estimates after a motion event is greater for RSS and SUB than for CIR and CSI. But all RF technologies are affected by motion to some degree. 

\begin{figure}[tb]
  \begin{center}
  $\vcenter{\hbox{\includegraphics[scale=0.45]{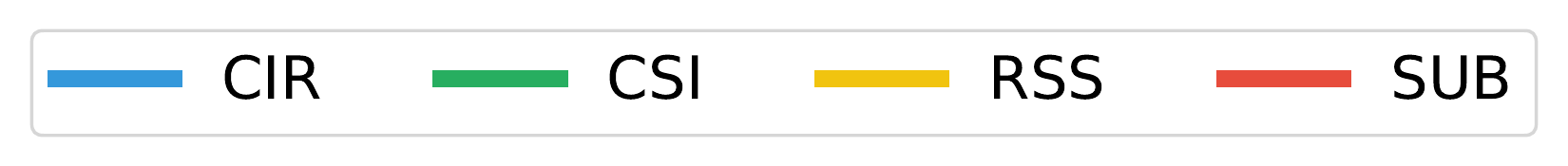}}}$
  $\vcenter{\hbox{\includegraphics[scale=0.40]{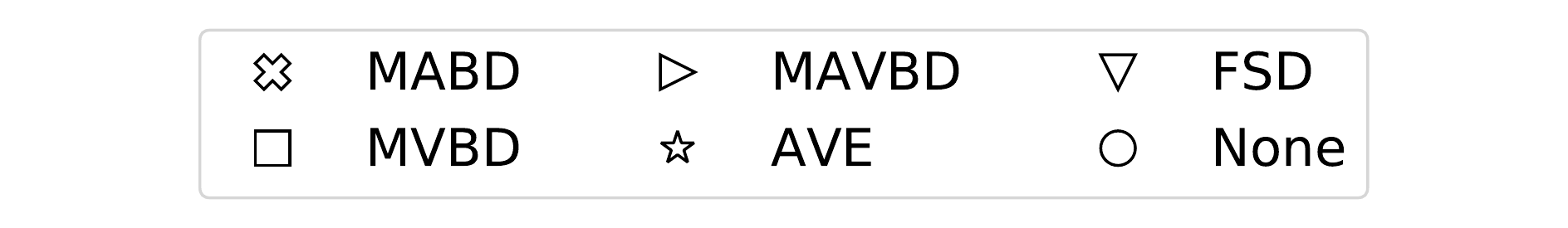}}}$
  $\vcenter{\hbox{\includegraphics[scale=0.45]{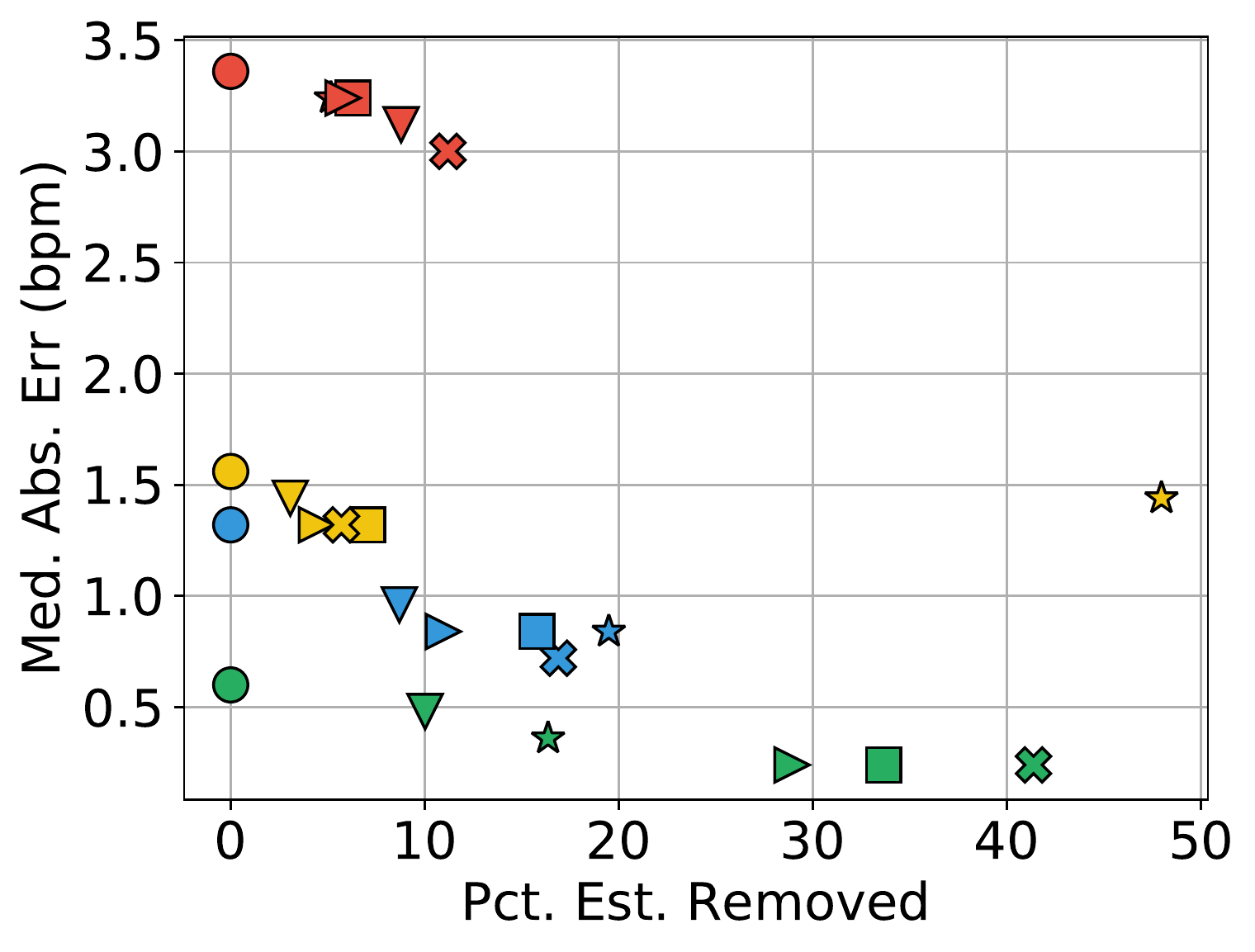}}}$
  \caption{The median $e$ and \% respiration rates ignored because of detected motion vs.\ motion detection method and RF technology.}
  \label{F:motion_err_compare}
  \end{center}
  \vspace{-1em}
\end{figure}

To negate the periods of large $e$ due to motion, we compare the five motion detection methods described in Section~\ref{S:motion_detection}. We enable stream select and use the PSD method for RR estimation. Each motion detection algorithm is evaluated one at a time. We compare the motion detection performance using the median $e$ achieved over all 20 studies and the percent of RR estimates that are ignored because of detected motion. These metrics provide a way to judge each algorithm's ability to remove the large RR estimate errors caused by motion and the ability to remove only estimates that were made during periods of motion. The results of this evaluation is shown in Fig.~\ref{F:motion_err_compare}.
 We observe that all of the motion detectors improve the median $e$ when compared to the case where no motion detection is used. The MABD tends to lower the median $e$ the most for each RF technology. However, the percent of RR estimates removed when using MABD tends to increase as well. From our observations, there is on average one motion event every 5~\si{\minute}. Motion events during sleep do not last any more than 30~\si{\second}. So over the course of an \SI{8}{\hour} study, there should be $10\%$ of the time when we are not estimating a respiratory rate. We use this percentage as a benchmark for the motion detectors. Additional sensing devices would have been needed to be attached to the patient during the sleep study to get a more accurate ground truth for motion. Removing less than $10\%$ of the estimates means that more RR estimates could be affected by motion and thus increase the median $e$. Alternatively, removing more than $10\%$ of the RR estimates means that we are not estimating RR rate during periods when there is no motion. 

With this tradeoff in mind, we observe that the choice of motion detector varies with the RF technology. For SUB, the MABD reduces the median $e$ by 0.4 bpm and removes $12\%$ of the RR estimates. For RSS, MVBD lowers the median $e$ by 0.3 bpm and removes the most RR estimates. For CIR, MAVDB lowers the median $e$ by 0.5 bpm and removes $11\%$ of the RR estimates. Lastly for CSI, FSD comes the closest to $10\%$ RR estimates removed and lowers the median $e$ by 0.1 bpm. We note that the thresholds we chose for each motion detection algorithm plays a role in the percent of RR estimates removed. Additional evaluation would need to be performed to show a more complete representation of how each motion detection algorithm could perform with different thresholds. However, for the purpose of this paper, it is sufficient to say that applying a motion detection algorithm can lower the median $e$ by removing RR estimates that were made during a period motion causes large $e$. No one motion detection method was definitively the best in terms of balancing reduce median $e$ and ignoring true periods of motion. Picking a computationally lightweight algorithm would consequently be an attractive option.

\subsection{Sleep Position's Effect on RR Estimation}
Up to this point, we have compared a variety of methods that have been developed to estimate RR. However, we found that none of the methods were able to address the high percentage of RR estimates with large errors. In the next two sections we investigate reasons why the estimates are so unreliable at times. 

\begin{figure}
  \begin{center}
  \includegraphics[scale=0.35]{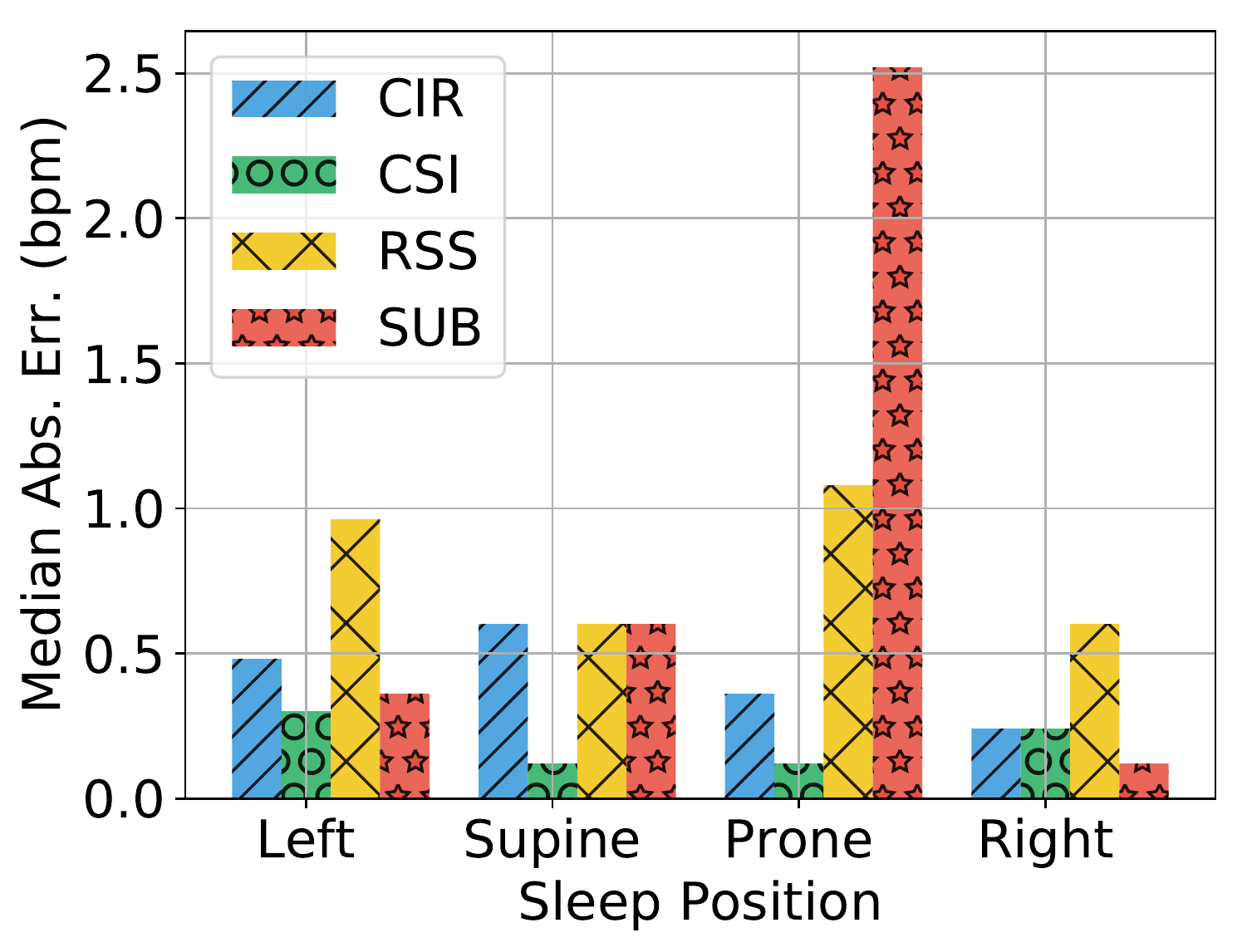}
  \caption{Median $e$ vs.\ RF technology when patient position is left, supine, prone, and right.}
  \label{F:err_afo_sleep_pos}
  \end{center}
  \vspace{-1em}
\end{figure}

One explanation for why the errors are so large for all RF technologies in the upper percentiles (as seen in Fig.~\ref{F:err_stream_select_PSD}) could be the patient's sleeping position. To test this explanation, we chose eight studies where all four RF technologies have average or better-than-average absolute errors compared to all twenty studies. During the sleep studies, technicians periodically annotate the patient's sleep position from their live video feed. We do not know, after the fact, the patient's true sleep position at every moment of the study. As an alternative, we mark times when the sleep position is annotated. We then record the time the first motion event occurred before and after the annotated sleep position timestamp. This way we know the patient was in a given sleep position for a given period of time. Motion events are found when all RF technologies visually show a significant change in mean and/or variance in their channel measurements. In all, we find 18 periods between 25 and 65 minutes long where the sleep position is known. The patient is sleeping on their left side for six of the periods, supine for five of the periods, prone for three of the periods, and on their right side for four of the periods. We plot the median $e$ during each of these periods and for each RF technology as shown in Fig.~\ref{F:err_afo_sleep_pos}.
We observe that in almost all cases, the median $e$ does not change considerably as a function of sleep position for a given RF technology. The median $e$ is 1.1 bpm or less. The exception is the SUB median $e$ when the patient was sleeping in the prone position. However, we only had three periods when the sleeping position was known to be prone. Thus we cannot find any strong evidence that RF breathing monitoring in general is impacted in any consistent manner by sleep position.

\subsection{RR Estimation Robustness}
Unlike the controlled and short studies conducted in prior research, the patients monitored during the sleep studies were not instructed to lie in a certain position for a certain amount of time. Veritably, patients would move during the study and could change their sleeping position at will. One reality with wireless RF respiratory monitoring is that even slight changes in a person's position will alter the fading characteristics of each stream of an RF device. When the multipath components add destructively at the receiver, the respiration signal will be hidden in noise or will be nonexistent. When this is the case for a large percentage of selected streams, RR estimation is very difficult.

\begin{figure}
  \begin{center}
  $\vcenter{\hbox{\includegraphics[scale=0.25]{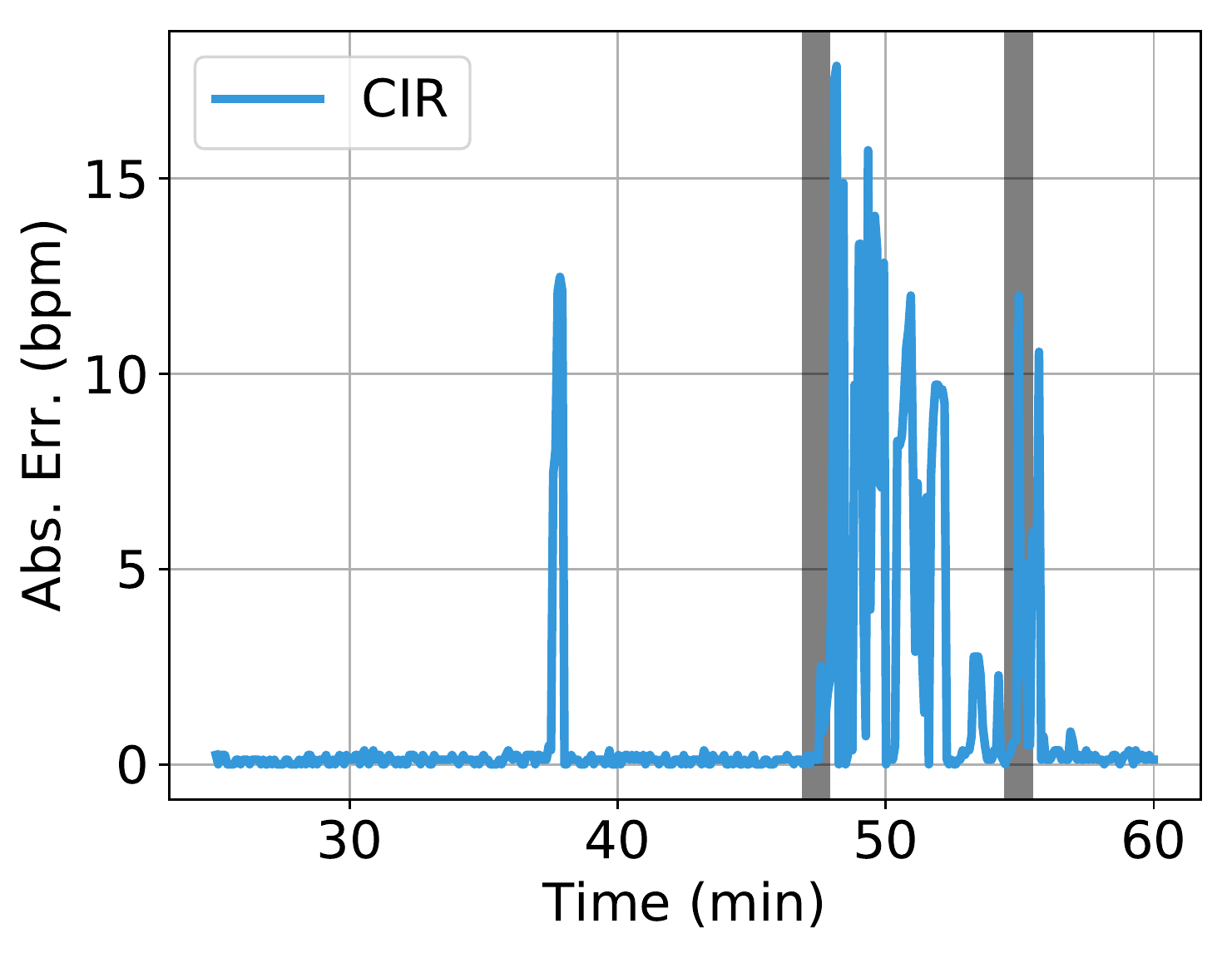}}}$\quad
  $\vcenter{\hbox{\includegraphics[scale=0.25]{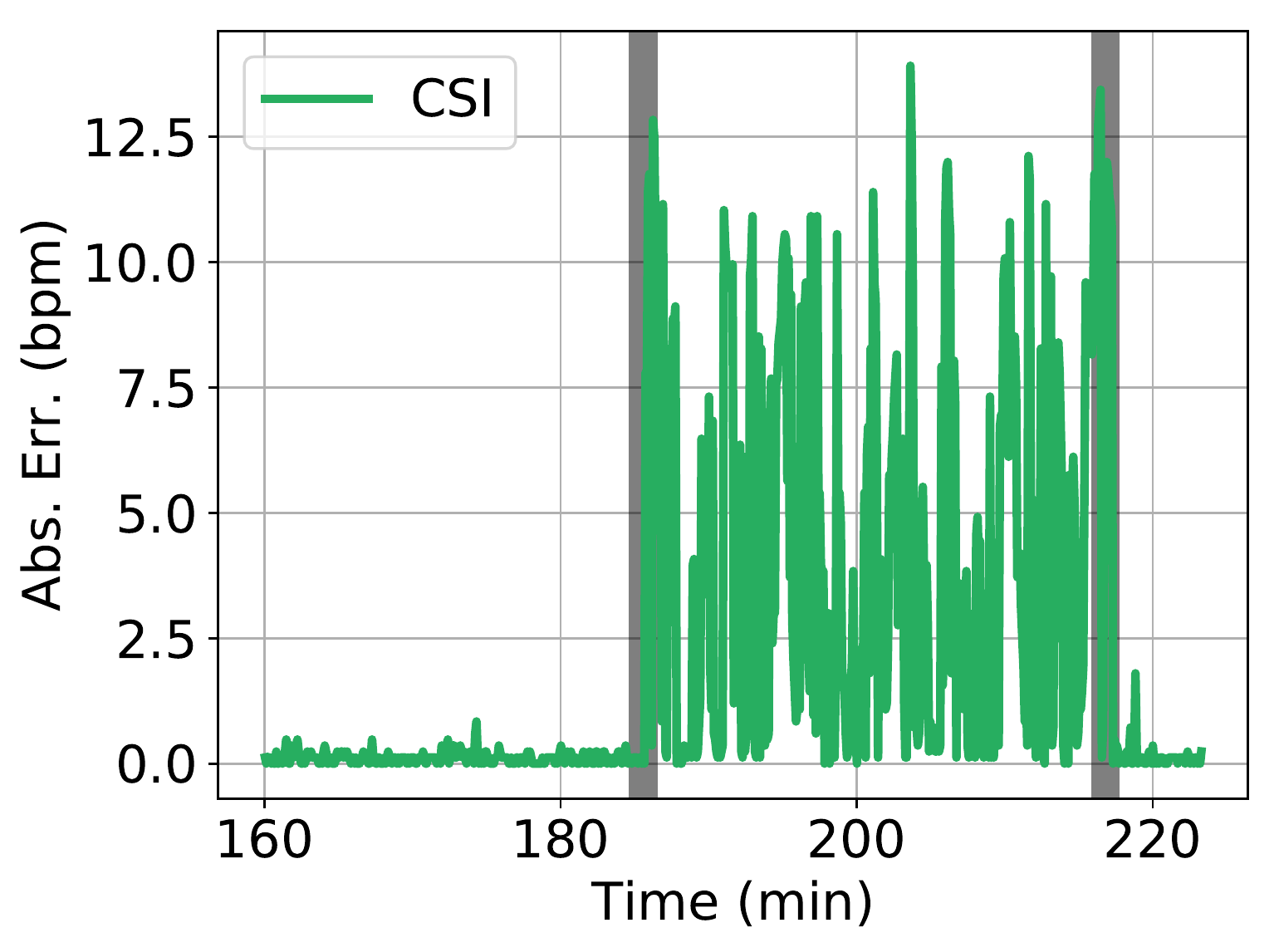}}}$
  $\vcenter{\hbox{\includegraphics[scale=0.25]{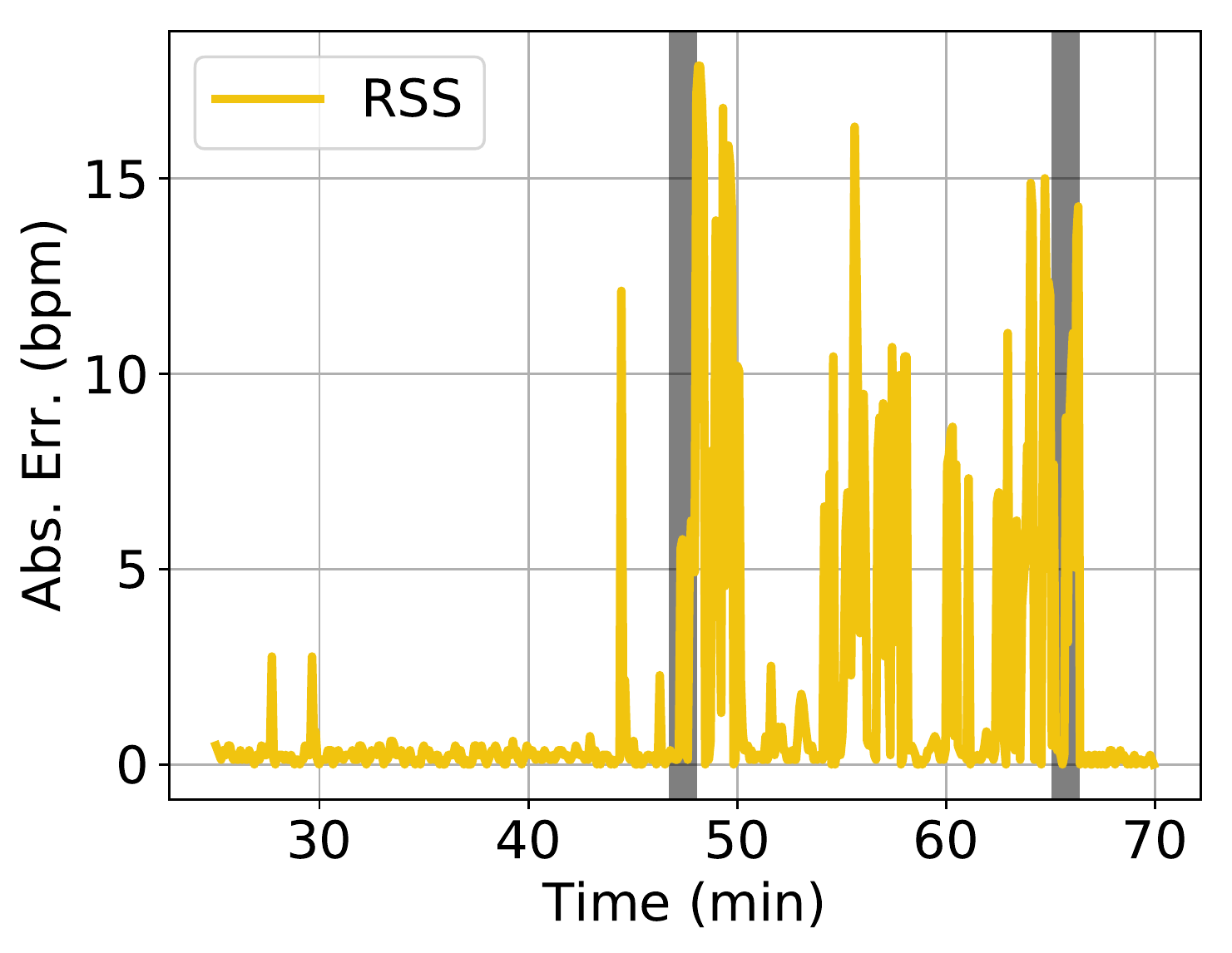}}}$\quad
  $\vcenter{\hbox{\includegraphics[scale=0.25]{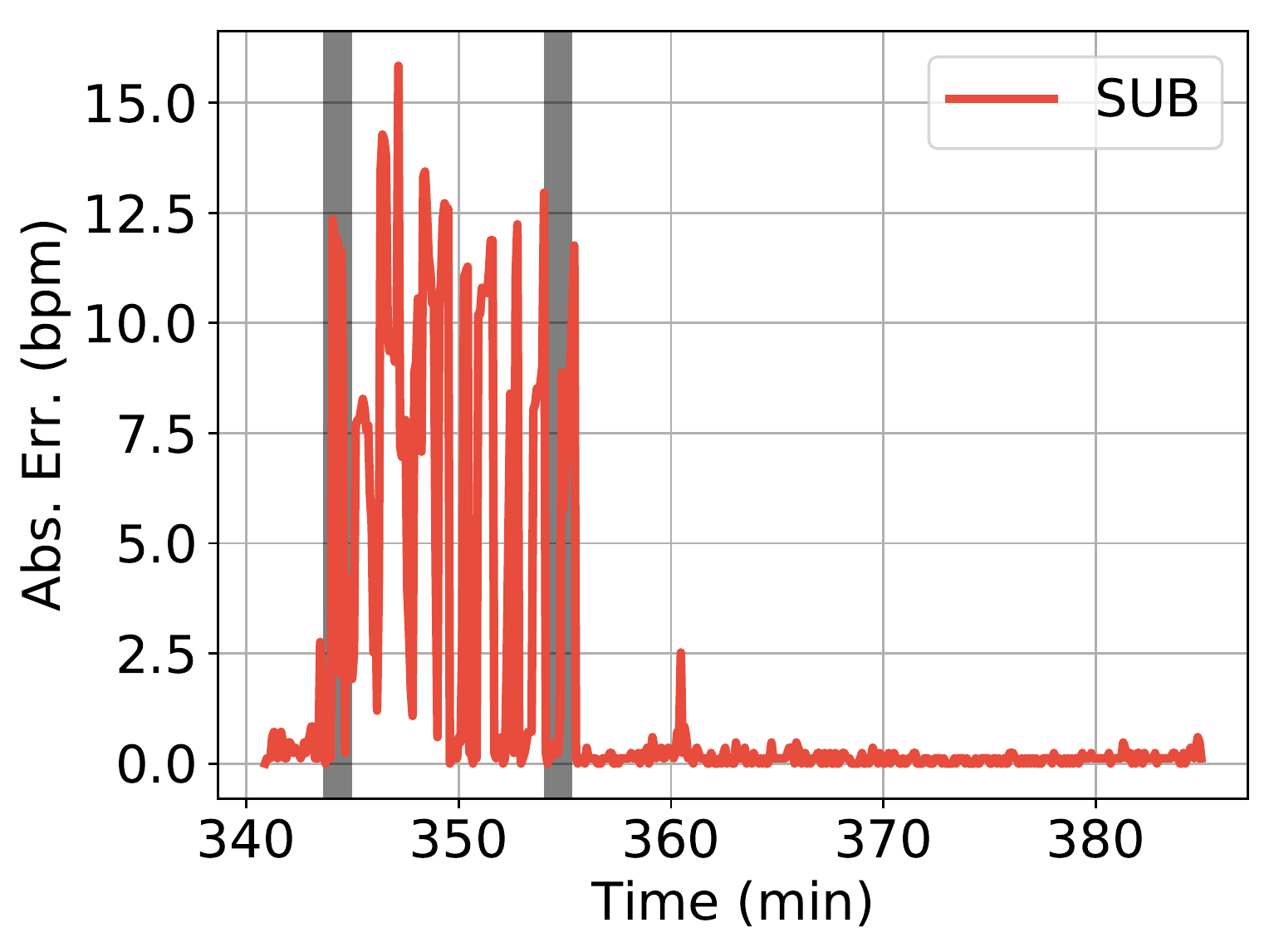}}}$
  \caption{The absolute error $e$ vs.\ method 
  during several minutes of different study periods. 
  The gray bands indicate two times when the patient moved during the period shown.}
  \label{F:gbg_rr_err}
  \end{center}
  \vspace{-1em}
\end{figure}

To demonstrate this reality, we plot different sections of time from a patient study when each RF device has periods of low then high RR estimation error that are related to periods of motion. For the plots shown in Fig.~\ref{F:gbg_rr_err}, we use stream selection, the PSD RR estimation method, but no motion detection.
 The gray bands in Fig.~\ref{F:gbg_rr_err} indicate the two times the patient moved during the window of errors shown. All four RF technologies have periods where the $e$ is much lower than 1 bpm. When the patient moves, the errors are higher and appear to be more uniformly distributed. The patient moves again after some time, and the error drops back down below 1 bpm error. The median absolute error before the first motion event, between the two motion events, and after the last motion event for each RF technology is summarized in Table~\ref{T:gbg_median_err}.
\begin{table}
\centering
\caption{Median $e$ (bpm) for each RF technology for errors shown in Fig.~\ref{F:gbg_rr_err}}
\label{T:gbg_median_err}
\begin{tabular}{c|cccc}
Motion Events & CIR & CSI & RSS & SUB \\ \hline \hline
Before & 0.12 & 0.12 & 0.24 & 0.24 \\
Between & 2.22 & 3.12 & 1.0 & 7.38 \\
After & 0.24 & 0.12 & 0.12 & 0.12
\end{tabular}
\vspace{-1em}
\end{table} We see that the median RR estimates for all RF technologies are well below 1 bpm error before and after the motion events. In fact, there is very little difference in estimation error when fading conditions are favorable for each RF technology. We also observe that every RF technology suffers from very unreliable RR estimates during some periods. When the fading conditions are not favorable, the median error increases for all RF technologies. This result is surprising for CSI and CIR. Despite having more diversity from using multiple antennas and/or operating with a large bandwidth, CSI and CIR are not able to reliably track RR at all times. CSI and CIR suffer just like the RSS-based  technologies from being in unfavorable fading conditions for periods of time.

An argument could be made that there is no difference in RR estimation among the technologies, conditioned on having favorable fading conditions. Consequently, COTS devices that only measure RSS on narrow channel bandwidths could be just as useful in respiration monitoring as a WiFi or UWB device that use much wider channel bandwidths. The difference is that CSI appears to be in a favorable fading condition the greatest amount of time. One question to ask is, is it possible to increase the amount of time that favorable fading exists? For example, achieving accurate RR estimates may be a matter of a new method that is designed to either select or weight streams based on their respiratory signal SNR. It is reasonable to assume that at least one stream would be able to reliably estimate RR. 
Stream selection may be the most fruitful path to explore since the sensitivity of the measurements to breathing ultimately dictates the quality of an respiratory rate estimation method.

Alternatively, RR estimation could be improved by adding one or two more transceivers somewhere around the person to increase the spatial diversity of the measurements. A link with poor fading characteristics could be ignored when an RF link with a different spatial orientation has more favorable fading. Although increasing the number of devices comes at a greater price-per-system cost and additional protocol development, the drawback would be worth the cost if the amount of time the RR estimate is reliable is increased.

%% file: future_conclusion.tex
\section{Conclusion}
In this paper, we developed a respiratory monitoring testbed containing four different RF devices to simulatenously measure the wireless channel. Included were WiFi devices that measured CSI, UWB radios that measured CIR, and two narrowband radios that measured 1-dB quantized and sub-dB quantized RSS. The testbed, placed in a sleep study room, collected channel measurements all night during twenty polysomnographies. The four RF technologies and several published methods that performed stream selection, motion detection, and RR estimation were then compared using the error in the RR estimate as a metric.

We found that CSI measurements resulted in the lowest median absolute error in RR estimates compared to CIR, RSS, and SUB. Yet, even with multiple antennas and/or large bandwidth, CSI and CIR failed to reliably track the respiratory rate over the long term just like the RSS and SUB which did not have the same points of diversity. In our results, we found that a PSD-based RR estimation performed better than an IBI approach based on the median absolute error achieved. In other tests, we found that there was no signficant difference in performance between motion detection algorithms. This study showed that many RF-based systems can perform respiratory monitoring equally reliably at times, but also that each system could be improved to overcome periods of time when the fading conditions are not favorable.

%% file: ack.tex
\section*{Acknowledgment}
\anontext{[removed for double-blind review]}{Research reported in this publication was supported in part by the National Institute on Drug Abuse of the National Institutes of Health under Award Number \#DA041960 and by the U. S. Army Research Office grant \#69215CS.  The content is solely the responsibility of the authors and does not necessarily represent the official views of the NIH or the ARO.}